\documentclass{aa}
\usepackage{graphics}
\usepackage{psfig}
\usepackage{lscape}
\newcommand{\ignore}[1]{}
\def\asec{{'' \hspace{-4pt} . \hspace{2pt} }}

\def\50mu{{< \hspace{-4pt} \mu_{50} \hspace{-4pt}>}}
\def\75mu{{< \hspace{-4pt} \mu_{75} \hspace{-4pt}>}}
\def\mmu{{< \hspace{-4pt} \mu_m \hspace{-4pt}>}}
\def\avmu{{< \hspace{-4pt} \mu \hspace{-4pt}>}}

\def\ksM{km\thinspace s$^{-1}$\thinspace  Mpc$^{-1}$}
\def\gs{\mathrel{\raise0.35ex\hbox{$\scriptstyle >$}\kern-0.6em
\lower0.40ex\hbox{{$\scriptstyle \sim$}}}}
\def\ls{\mathrel{\raise0.35ex\hbox{$\scriptstyle <$}\kern-0.6em
\lower0.40ex\hbox{{$\scriptstyle \sim$}}}}
\def\et{\hbox{et al.}$\,$}
\newenvironment{changemargin}[4]{%
    \begin{list}{}{%
     \setlength{\topsep}{0pt}%
     \setlength{\leftmargin}{#1}%
     \setlength{\rightmargin}{#2}%
     \setlength{\topmargin}{#3}%
     \setlength{\textheight}{#4}%
     \setlength{\listparindent}{\parindent}%
     \setlength{\itemindent}{\parindent}%
     \setlength{\parsep}{\parskip}%
    }%
   \item[]}{\end{list}}
\begin{document}

%
   \title{The scaling relations of early--type galaxies in clusters}

   \subtitle{I. Surface photometry in seven nearby clusters}

   \author{
           G. Fasano\inst{1}
           \and D. Bettoni\inst{1}
           \and M. D'Onofrio\inst{2}
           \and P. Kj{\ae}rgaard\inst{3}
           \and M. Moles\inst{4}
          }

   \offprints{
              G. Fasano
             }

   \institute{
              Osservatorio Astronomico di Padova,
              Vicolo Osservatorio 5, I-35122 Padova,
              email: fasano@pd.astro.it, bettoni@pd.astro.it
         \and
              Dipartimento di Astronomia dell'Universit\`a di Padova,
              Vicolo Osservatorio 2, I-35122 Padova,
              email: donofrio@pd.astro.it
         \and
              Copenhagen University Observatory. The Niels Bohr Institute for
              Astronomy, Physics and Geophysics\\ Juliane Maries Vej 30, DK-2100
              Copenhagen,
              email: per@astro.ku.dk
         \and
              Instituto de Matem\'{a}ticas y F\'{\i}sica Fundamental, CSIC,
              C/ Serrano 113B, 28006 Madrid, Spain,
              email: moles@imaff.cfmac.csic.es
             }

   \date{
         Received December 2000
        }

\abstract{
This is the first paper of a series investigating the scaling
relations of early--type galaxies in clusters. Here we illustrate the
multi--band imagery and the image reduction and calibration procedures
relative to the whole sample of 22 clusters at 0.05 $\ls$ z $\ls$
0.25.  We also present the detailed surface photometry of 312
early--type galaxies in 7 clusters in the first redshift bin,
$\approx$0.025--0.075. We give for each galaxy the complete set of
luminosity and geometrical profiles, and and a number of global,
photometric and morphological parameters. They have been evaluated
taking into account the effects of seeing. Internal consistency checks
and comparisons with data in the literature confirm the quality of our
analysis. These data, together with the spectroscopic ones presented
in the second paper of the series, will provide the local calibration
of the scaling relations.
}

   \maketitle

%

\section{Introduction}

Clusters of galaxies are the most massive, yet dynamically bound,
known entities in the Universe. The identification of properties that
could be universal would make them tracers of cosmic evolution since
they can be detected at large distances. Indeed, the scaling relations
satisfied by the global parameters of the early type galaxies, the
dominant population in the densest parts of clusters, have become a
powerful tool to elucidate the nature of the processes of formation
and evolution of galaxies, as well as to perform different
cosmological tests.

In general, assuming that early type (E or S0) galaxies of mass M and
luminosity L, are in virial equilibrium, and that they all are
homologous systems, it follows that some relation between size,
surface brightness and velocity dispersion would be expected, provided
that the M/L ratio is a function of the same variables. It is now well
established that such a relation, the so called Fundamental Plane (FP
hereafter), does exist (Dressler et al.~1987, Djorgovski \&
Davis~1987, J{\o}rgensen, Franx and Kj{\ae}rgaard~1995). It has the
form R$_e \propto\sigma^{\alpha}\times$ I$_{e}^{\beta}$, where R$_e$
is the effective radius, $\sigma$ the velocity dispersion, and
$<$I$>_{e}$ the mean effective surface brightness.

The existence of the FP implies that the main physical relation
governing the properties of E and S0 galaxies is just the virial
condition (Faber et al. 1987). The precise form of the FP relation
depends on different factors, such as the lack of exact homology of
the early type galaxies in different contexts (i.e., differences in
the luminosity profiles or in the dynamical structure: Caon et
al.~1993, Graham et al.~1996, Ciotti et al.~1996) and the dependence
of the M/L {\sl versus} M relation on wavelength (Pahre and
Djorgovski~1995; Pahre, Djorgovski, \& de Carvalho~1998; Pierini et
al.~2000; Mobasher et al.~1999). Moreover, it should be established
which galaxies (i.e.: their luminosity range) are appropriate to build
the FP. Given the number of factors that could contribute to the
scatter of the FP, its small amplitude is remarkable. It is that small
scatter which makes the FP an accurate distance indicator. The
Kormendy relation, the projection of the FP onto the R$_e$--I$_{e}$
plane, has substantially more scatter.  However, if part of that
scatter could be understood in terms other than the velocity
dispersion, its suitability for cosmological analysis could be
reinforced.  As discussed by Kj{\ae}rgaard, J{\o}rgensen and Moles
(1993; KJM hereafter), the rapidly increasing difficulty, for
increasing cluster redshift, to obtain the velocity dispersion of a
sizeable number of galaxies, results in defining the FP with rather
limited samples, implying an increasing scatter.

In all the studies it is assumed that the Fundamental Plane, as well
as its projections, the Kormendy relation in particular, are
universal, in the sense of presenting the same coefficients
everywhere, independently of any local property or even of the
redshift. Indeed, it is the assumption of universality of the FP that
could make it an appropriate tool in cosmology, e.g. in performing the
Tolman test (KJM; Pahre, Djorgovski, \& de Carvalho~1996; Moles et
al.~1998), or to assess the evolution of M/L with z (Bender, Burstein,
\& Faber ~1992; Guzm\'an, Lucey, \& Bower 1993; van Dokkum and
Franx~1996; Kelson et al.~1997; J{\o}rgensen, \& Hjorth~1997; Bender et 
al.~1998; Ziegler et al.~1999; J{\o}rgensen et al.~1999;
Kelson et al.~2000). This assumption, however, has to be empirically verified. 

J{\o}rgensen, Franx, \& Kj{\ae}rgaard (1996; hereafter JFK) have
discussed the problem and presented a study of early type galaxies in
ten clusters, finding that the distribution of their structural
parameters can be fitted with a unique set of coefficients for all the
clusters. However, as indicated by JFK, given the small sample size,
in some cases, some variations of the order of 10\% in the
coefficients cannot be excluded. Their sample included clusters with a
wide range of richness and regularity, with z$\leq$0.038. Work at
higher redshift (van Dokkum and Franx 1996) shows that the data can be
consitently fitted with similar coefficients, but cannot be considered
as definitive due to the small number of galaxies used to define the
relations.

The program presented in KJM addresses the same question of the
universality of the scaling relations, trying to separate the cosmic
variance at a given redshift, from cosmic evolution effects in a
systematic way. Our work includes galaxy clusters with a more
restricted range of properties than in JFK, in the redshift range
0.035$\leq$ z $\leq$0.28, in 5 redshift steps. The comparison of the
results for galaxy clusters at a similar redshift will allow to
control the effects of local aspects of the evolution on the FP
coefficients. Going to significantly higher redshift that JFK, we'll
be able to test the behaviour of the scaling relations with z. We
stress here that we not only consider in our program the FP relation,
but, for the resons pointed out before and discussed by KJM, we'll
analyze the properties and behaviour of the Kormendy relation as well.

Another interesting aspect we want to study is the characterization of
the family of galaxies that define and satisfy the scaling relations,
an aspect not yet understood. Regarding the Kormendy relation, it is
known that it is satisfied by only a fraction of early type galaxies
in the (I$_e$-r$_e$) plane (Capaccioli, Caon, \& D'onofrio 1992). It
is one of our goals to analyze where the borderline of these two
families lies, and the reason of such a dichotomy.  Concerning the FP,
even the deepest studies to now (J{\o}rgensen 1999) only include the
brightest end of the lumnosity function, as the measurement of the
velocity dispersion of fainter early type galaxies needs of important
amounts of observing time with very large telescopes.

The data we collected, following the scheme proposed by KJM, include
surface photometry and spectroscopy (intermediate and low resolution)
of a sizeable number of E and S0 galaxies in each cluster, to get not
only the parameters $<$I$>_e$, R$_e$ and $\sigma$, but also a measure
of the K-effect for each galaxy, together with their spectral energy
distributions and some spectral indicators.

In the present paper, which is the first step to achieve such a
program, we give an overview of all the photometric observations
relative to the project, illustrating the sample selection and the
observing strategy, describing the reduction and calibration of the
data and discussing their quality (Section~2). We present here the
detailed surface photometry of the galaxies, down to $M_r\sim
-$18.0~mag (H$_0$ = 75~\ksM), for 7 nearby (z$\ls$0.075) clusters of
the sample (Section~3), describing the procedure we used to extract
the global photometric and morphological parameters of galaxies
(Sections~4,5). Finally we perform internal and external comparisons
to check the reliability of our analysis (Section~6).  The results on
more distant clusters and the properties of the scaling relations will
be given in following papers. Some preliminary results were
anticipated in Fasano et al. (1997), and in Moles et al. (1998), where
we have used the Kormendy relation to perform the Tolman surface
brightness test. In Fasano et al.(2000) we have also presented the
analysis of the morphological content (the E/S0/S fractions) in
clusters up to z $\sim$ 0.5, including some of our sample. Through
this paper we assume $H_0$=75~\ksM and $q_0$=0.1.

\section{The global sample: observations and data reduction}

\subsection{Sample selection and cluster coverage}

Table~\ref{ClTSam} lists the clusters in our sample and the observing
runs in which they were observed (see next subsection).  The clusters
have been grouped in five redshift bins with step $\sim$0.05, from
z$\sim$0.025 up to z$\sim$0.25. They were selected from the catalog by
Abell et al.~(1989) to be representative of massive, apparently
relaxed systems, not too different from Coma, a regular cluster for
which an important amount of relevant data is available. Thus we
selected clusters of intermediate richness class, with Bautz-Morgan
(1970) type II, II-III or III. Types I and I-II were excluded to avoid
clusters dominated by a central, big cD galaxy. We also excluded those
clusters with Rood-Sastry (1971) class L or I, to maximize the
probability of dealing with virialized systems. Finally, we restricted
the sample to rather high galactic latitudes (most have
$|b|\geq$36$^\circ$) in order to avoid problems with the
extinction. The 22 clusters we actually observed (see
Table~\ref{ClTSam}) are from the resulting list, the final selection
being a matter of opportunity. We notice that there is an exception to
our selection criteria, namely A2670, a cD dominated cluster, BM class
I-II.

   \begin{table*}
      \caption[]{The global sample of clusters}
         \label{ClTSam}
      \[
         \begin{array}{llllllllccc}
            \hline\noalign{\smallskip}

\multicolumn{1}{c}{\rm Name}&
\multicolumn{3}{c}{\alpha (2000)^{(1)}}&
\multicolumn{3}{c}{\delta (2000)^{(1)}}&
\multicolumn{1}{c}{z}&
\multicolumn{1}{c}{\rm BM type^{(2)}}&
\multicolumn{1}{c}{\rm RS type^{(3)}}&
\multicolumn{1}{c}{\rm runs}\\
\hline\noalign{\smallskip}
{\rm A~2151} & 16^{h} & 05^{m} & 15^{s} & +17^{\circ} &
44^{\prime} & 55^{\prime\prime} & 0.0365 & {\rm III} & F & 2,3\\
{\rm A~119} & 00 & 56 &  21 & -01 & 15 & 47 & 0.0439 & {\rm  II-III} & C & 1,5,7\\
{\rm A~1983} & 14 & 52 & 44 & +16 & 44 & 46 & 0.0456 & {\rm III} & F & 3\\
{\rm DC~2103} & 21& 08 & 37 & -39 & 51 & 19 & 0.0527 & -   & - & 4\\
{\rm A~3125} & 03 & 27 & 22 & -53 & 30 & 38 & 0.0593 & {\rm III} & - & 5\\
{\rm A~1069} & 10 & 39 & 54 & -08 & 36 & 40 & 0.0630 & {\rm III} & F & 5\\
{\rm A~2670} & 23 & 54 & 10 & -10 & 24 & 18 & 0.0761 & {\rm I-II} & cD & 7\\
\hline
{\rm A~3330} & 05 & 14 & 42 & -49 & 03 & 00 & 0.0910 & {\rm II} & - & 1,5\\
{\rm A~2048} & 15 & 15 & 18 & +04 & 22 & 00 & 0.0945 & {\rm III} & C & 3\\
{\rm A~98}   & 00 & 46 & 24 & +20 & 29 & 00 & 0.1015 & {\rm II-III} & F & 7\\
{\rm A~216}  & 01 & 36 & 42 & -06 & 24 & 00 & 0.1158 & {\rm II-III} & C & 7\\
{\rm A~389}  & 02 & 51 & 18 & -24 & 54 & 00 & 0.1160 & {\rm II} & F & 5\\
\hline
{\rm A~951}  & 10 & 13 & 54 & +34 & 43 & 00 & 0.1430 & {\rm III} & B & 6\\
{\rm A~2235} & 16 & 55 & 00 & +40 & 01 & 00 & 0.1511 & {\rm III} & F & 7\\
{\rm A~1979} & 14 & 51 & 00 & +31 & 16 & 00 & 0.1687 & {\rm II-III} & F & 3\\
\hline
{\rm A~2658} & 23 & 45 & 00 & -12 & 18 & 00 & 0.1850 & {\rm III} & F & 3\\
{\rm A~2192} & 16 & 26 & 36 & +42 & 40 & 00 & 0.1868 & {\rm II-III} & F & 2,3\\
{\rm A~1643} & 12 & 55 & 54 & +44 & 04 & 00 & 0.1980 & {\rm III} & B & 2\\
\hline
{\rm A~2111} & 15 & 39 & 36 & +34 & 24 & 00 & 0.2290 & {\rm II-III} & C & 2\\
{\rm A~2125} & 15 & 35 & 54 & +70 & 03 & 00 & 0.2465 & {\rm II-III} & B & 3\\
{\rm A~1952} & 14 & 41 & 06 & +28 & 38 & 00 & 0.2480 & {\rm III} & C & 3\\
{\rm A~1878} & 14 & 12 & 48 & +29 & 12 & 00 & 0.2540 & {\rm II} & C & 2\\

            \noalign{\smallskip}\hline

\noalign{\medskip}

\multicolumn{11}{l}{\rm ^{(1)}~NED~coordinate} \\
\multicolumn{11}{l}{\rm ^{(2)}~Bautz-Morgan~(1970)~Type} \\
\multicolumn{11}{l}{\rm ^{(3)}~Rood-Sastry~(1971)~Class} \\

         \end{array}
      \]
   \end{table*}

The choice of the cluster fields to be observed in the framework of
our global program is a more delicate question. The ideal approach
would be to cover the whole cluster within $\sim$1 Abell
diameter. Indeed, for the determination of the FP, measuring the
$\sigma$ for substantially more than the $\sim$15-20 brightest early
type galaxies in each cluster is too consuming in telescope time, and
would require a separate observing program. On the other hand, to
tackle the question of the family of galaxies actually defining the
scaling relations, in particular the Kormendy relation, it would be
necessary to have a fairly complete sample reaching some (faint)
absolute magnitude limit. This is a program now under way, but for the
time being we tried to optimize the effort making some
compromises. Basically, for the nearest clusters (see Section~3) only
individual targets were selected from the catalog by Dressler (1980)
-- the brightest E and S0 galaxies in his lists. Consequently only the
galaxies entering the fields of those individual targets were recorded
and measured (see Table~\ref{Fields}). For more distant clusters the
coverage approaches the ideal, since a substantial fraction of the
whole cluster was sampled. The final data collection includes several
dozens of galaxies per cluster with accurate surface photometry.

\subsection{Observations and basic reductions}

The observations for the whole sample of 22 clusters were collected
from Dec.~1994 to Aug.~1998 with the Nordic Optical Telescope (NOT, La
Palma) and with the 1.5m Danish telescope at La Silla (Chile). Images
have been obtained in two or three bands [Gunn~$r$ ($r$), Bessel~V
($V$) and Bessel~B ($B$)] to provide the color term of each
object. Moreover, as a general strategy, at least two exposures for
each field in each filter were usually taken, allowing us to clean-up
the combined images for cosmic-rays.


   \begin{table*}
      \caption[]{Observing Runs}
         \label{ObsLog}
      \[
         \begin{array}{llccccccc}
            \hline\noalign{\smallskip}

  \multicolumn{1}{l}{\rm Run}
& \multicolumn{1}{c}{\rm Date }
& \multicolumn{1}{c}{\rm Telescope}
& \multicolumn{1}{c}{\rm Camera}
& \multicolumn{1}{c}{\rm CCD}
& \multicolumn{1}{c}{\rm read noise}
& \multicolumn{1}{c}{\rm gain}
& \multicolumn{1}{c}{\rm Pixel Scale}
& \multicolumn{1}{c}{\rm F.O.V.} \\
 & & & & & (e) & (e/{\rm ADU}) & (^{\prime\prime}/pixel) & ^{arcmin\times arcmin} \\
\hline
 & & & & & & & & \\
1 & {\rm Dec.~4-8~1994} & {\rm 1.5 Danish} & {\rm DFOSC} & {\rm Thompson} & 3.70
& 2.00 & 0.510 & 8.7\times 8.7 \\ 2 & {\rm May~24-27~1995} & {\rm NOT} & {\rm
StanCam} & {\rm Tektronik} & 6.36 & 1.69 & 0.176 & 3\times 3 \\ 3 & {\rm
Jun.~23-26~1995} & {\rm NOT} & {\rm StanCam} & {\rm Tektronik} & 6.36 & 1.69 &
0.176 & 3\times 3 \\ 4 & {\rm Sep.~19~1995} & {\rm 1.5 Danish} & {\rm DFOSC} &
{\rm Loral} & 4.90 & 1.31 & 0.390 & 13.4\times 13.4 \\ 5 & {\rm Jan.~1-8~1997} &
{\rm 1.5 Danish} & {\rm DFOSC} & {\rm Loral} & 4.90 & 1.31 & 0.390 & 13.4\times
13.4 \\ 6 & {\rm Feb.~27-28~1997} & {\rm NOT} & {\rm ALFOSC} & {\rm Ford-Loral}
& 6.50 & 1.02 & 0.189 & 6.5\times 6.5 \\ 7 & {\rm Aug.~29~1998} & {\rm NOT} &
{\rm ALFOSC} & {\rm Ford-Loral} & 6.50 & 1.02 & 0.189 & 6.5\times 6.5\\

            \noalign{\smallskip}\hline
         \end{array}
      \]
   \end{table*}


In Table~\ref{ObsLog} we report the log of observations, together with
the basic information about the instrumentation.
In each observing run, besides the cluster fields, a number of
nearby standard galaxies were imaged in the above mentioned bands,
thus allowing internal and external check of the surface photometry
(see Section 6).

Dark counts with different exposure times were obtained in each run
and turned out to be negligible. For the bias, several frames were
recorded every night to check the stability of the corresponding frame
structure. We also obtained, every night, several flat field (FF),
twilight sky exposures which, after comparison, were used to derive an
average frame. Apart from run~\#5, the quotient frames obtained using
FFs relative to different nights of the same run, turned out to be
almost flat, allowing us to produce a grand-averaged, high S/N ratio
FF to be used for the whole run.

%
%
%
%

The typical uncertainties associated with the reduction procedures are
of few tenths of ADU (0.1$-$0.2) for the bias removal and of few
thousandths of the background (0.001$-$0.006) for the flat
fielding. These uncertainties, together with those relative to the
photometric calibration (see Table~\ref{CalCoef}), will be used in
Section~5 to evaluate the expected errors in the global photometric
and morphological parameters.

During run~\#1 we used DFOSC during its testing phase, paying this
choice with some instability of the acquisition system. The most
serious drawback was the unreliability of the header content,
including the exposure time and the zenith distance. In the next
subsection we will report on the problems caused by this fact in the
calibration procedure.

Runs~\#2 and \#3 were characterized by exceptional weather
conditions. All the nights were photometric and the seeing turned out
to be permanently and largely below one arcsecond, with a minimum
value around 0$\asec$45 (run~\#2, May~24 1995).

During the run~\#5 the seeing was mediocre (1\farcs3--1\farcs5). We
had six photometric nights grouped in two intervals (Jan.~1--3 and
6--8), which turned out to be slightly different as far as the
reduction and calibration parameters are concerned (see
Table~\ref{CalCoef}).

\subsection{Calibration to standard passbands}

During each observing night, several (from 30 to 60) standard stars
from Landolt (1992), J{\o}rgensen (1994) and Montgomery et al (1993),
were observed at different zenith distances. In some cases the
standard stars exposures were slightly defocused to avoid saturation
of the brightest stars. The radius for the aperture photometry was set
to 3.5$\times$FWHM, where the FWHM includes the possible effect of
defocusing.

The coefficients of the relations between instrumental magnitudes in
the $i^{th}$ band, $m_i^{raw}$=$-$2.5~log(counts/sec), and the
standard systems have been computed adopting general expressions of
the form:

\begin{equation}
  K_i = m_i^{std} - m_i^{raw} = C_0^i + C_c^i(ij)\times col^{std}_{ij}
+ C_z^i\times sec(z_d)
\end{equation}

\noindent
were $m_i^{std}$ is the standard magnitude in the $i^{th}$ band,
$col_{ij}^{std}$=($m_j^{std}-m_i^{std}$) is the color of the object in
the standard system, $z_d$ is the zenith distance and $C_0^i$,
$C_c^i(ij)$, $C_z^i$ are the night constant, the color coefficient and
the extinction coefficient, respectively.

The coefficients $C_0^i$ and $C_z^i$ depend on the observing
conditions (basically on the site and on the night), whereas the color
coefficient $C_c^i(ij)$ should only depend on the filter ($i$) and on
the color ($ij$) since it indicate how well the instrumental response
(telescope$+$filter$+$detector) match the adopted standard system.

Since in each given run we found similar extinction coefficients for
the nights when we did photometry, in order to derive the calibration
coefficients we adopted the following multistep procedure:

(i) for each run and filter, the instrumental magnitudes of individual
stars observed at different zenith distances were
compared with the corresponding magnitudes in the standard system in
order to determine a `run-averaged' extinction coefficient;
(ii) for each night and for each filter, the extinction coefficients 
were used to determine the night zero-point offsets;
(iii) for each filter, the 'zero--airmass' calibration coefficients,
reduced to a common offset, were then correlated with the standard
colors to get the color coefficients and the 'common offset' zero points;
(iv) for each night and for each filter, the final zero point was
obtained by adding the night offset to the corresponding 'common offset' 
zero point.


   \begin{table*}
      \caption[]{Calibration coefficients and errors in the $r$ band}
         \label{CalCoef}
      \[
         \begin{array}{ccccccc}
            \hline\noalign{\smallskip}

  \multicolumn{1}{c}{\rm Run}
& \multicolumn{1}{c}{\rm Night(s)}
& \multicolumn{1}{c}{\rm Color}
& \multicolumn{1}{c}{C_0}
& \multicolumn{1}{c}{C_c}
& \multicolumn{1}{c}{C_z}
& \multicolumn{1}{c}{\rm r.m.s.} \\
\hline
1 & {\rm Dec.~4}  & (B-r) & 22.802\pm 0.019 & 0.110\pm 0.018 & -0.087\pm 0.023 & 0.058 \\
1 & {\rm Dec.~4}  & (V-r) & 22.866\pm 0.015 & 0.342\pm 0.024 & -0.087\pm 0.023 & 0.060 \\
\hline
2 & {\rm May~24}  & (B-r) & 24.737\pm 0.007 & 0.085\pm 0.003 & -0.128\pm 0.013 & 0.021 \\ 
2 & {\rm May~25}  & (B-r) & 24.704\pm 0.004 & 0.085\pm 0.003 & -0.128\pm 0.013 & 0.022 \\ 
2 & {\rm May~26}  & (B-r) & 24.738\pm 0.005 & 0.085\pm 0.003 & -0.128\pm 0.013 & 0.027 \\ 
2 & {\rm May~27}  & (B-r) & 24.755\pm 0.006 & 0.085\pm 0.003 & -0.128\pm 0.013 & 0.031 \\
\hline
3 & {\rm June~23} & (B-r) & 25.123\pm 0.015 & 0.117\pm 0.005 & -0.088\pm 0.005 & 0.022 \\
3 & {\rm June~24} & (B-r) & 25.123\pm 0.009 & 0.117\pm 0.005 & -0.088\pm 0.005 & 0.030 \\ 
3 & {\rm June~25} & (B-r) & 25.111\pm 0.005 & 0.117\pm 0.005 & -0.088\pm 0.005 & 0.023 \\ 
3 & {\rm June~26} & (B-r) & 25.121\pm 0.008 & 0.117\pm 0.005 & -0.088\pm 0.005 & 0.036 \\
\hline
4 & {\rm Sep.~19} & (V-r) & 21.357\pm 0.021 & 0.467\pm 0.089 & -0.081\pm 0.003 & 0.020 \\
\hline
5 & {\rm Jan.~1-3}& (B-r) & 23.619\pm 0.003 & 0.075\pm 0.003 & -0.105\pm 0.030 & 0.015 \\
5 & {\rm Jan.~6}  & (B-r) & 23.569\pm 0.001 & 0.075\pm 0.003 & -0.105\pm 0.030 & 0.015 \\
5 & {\rm Jan.~7}  & (B-r) & 23.631\pm 0.001 & 0.075\pm 0.003 & -0.105\pm 0.030 & 0.015 \\ 
5 & {\rm Jan.~8}  & (B-r) & 23.663\pm 0.001 & 0.075\pm 0.003 & -0.105\pm 0.030 & 0.015 \\ 
5 & {\rm Jan.~1-3}& (V-r) & 23.696\pm 0.001 & 0.187\pm 0.008 & -0.105\pm 0.030 & 0.014 \\ 
5 & {\rm Jan.~6}  & (V-r) & 23.646\pm 0.001 & 0.187\pm 0.008 & -0.105\pm 0.030 & 0.014 \\ 
5 & {\rm Jan.~7}  & (V-r) & 23.708\pm 0.001 & 0.187\pm 0.008 & -0.105\pm 0.030 & 0.014 \\ 
5 & {\rm Jan.~8}  & (V-r) & 23.740\pm 0.001 & 0.187\pm 0.008 & -0.105\pm 0.030 & 0.014 \\
\hline
6 & {\rm Feb.~27} & (B-r) & 25.285\pm 0.005 & 0.074\pm 0.004 & -0.084\pm 0.005 & 0.021 \\
6 & {\rm Feb.~28} & (B-r) & 25.321\pm 0.005 & 0.074\pm 0.004 & -0.084\pm 0.005 & 0.022 \\
\hline
7 & {\rm Aug.~29} & (B-r) & 25.264\pm 0.009 & 0.124\pm 0.008 & -0.090\pm 0.003 & 0.027 \\

            \noalign{\smallskip}\hline
         \end{array}
      \]
   \end{table*}


Since we did $r$ band observations in all runs, we report in
Table~\ref{CalCoef} the best-fit values of the coefficients $K_r$ for
each observing night, together with the relative uncertainties and the
final r.m.s.

A special case is that of the run~\#1 due to the incompleteness of the
header information (see Section 2.2). The large uncertainties relative
to this run in Table~\ref{CalCoef} reflect these calibration
problems. For the time being we present here just the results from a
field which overlaps with other field observed in a different run. The
complete photometry of run~\#1 will be presented separately in a
forthcoming paper.

\section{The sample of early-type galaxies in nearby clusters}

In the following we will concentrate on the surface photometry of
early type galaxies in the seven clusters of Table~\ref{ClTSam}
belonging to the first redshift bin (z$\ls$0.075). The results of the
present analysis will be used in a forthcoming paper to investigate
the properties of the galaxy scaling relations in nearby clusters (see
Section~1).

In column~1 of Table~\ref{NearClu} we list the clusters ordered by
increasing redshift, while columns~2 to 4 report, for each cluster,
the galactic extinction given by Schlegel et al.~(1998; see J\o
rgensen et al.~1994 for conversion to the $r$ band).


   \begin{table*}
      \caption[]{The nearby cluster sample}
         \label{NearClu}
      \[
         \begin{array}{lccccclccc}
            \hline\noalign{\smallskip}

 {\rm Cluster} & A_B^{(1)} & A_V^{(1)} & A_r^{(1)} & r_{lim}^{(2)} &
  N_{gal}^{(3)} & n_z^{(4)} & n_{dz}^{(5)} & n_{red}^{(6)} & n_{field}^{(7)} \\
\hline
{\rm A~2151}  & 0.205 & 0.158 & 0.128 & 17.86 & 37 & 23 & 2 & 3 & 1 \\
{\rm A~119}   & 0.167 & 0.128 & 0.104 & 18.26 & 57 & 27 & 2 & 1 & 4 \\
{\rm A~1983}  & 0.114 & 0.088 & 0.071 & 18.35 & 46 & 21 & 3 & 12 & 1 \\
{\rm DC~2103} & 0.176 & 0.130 & 0.110 & 18.67 & 49 & 20(15) & 7 & 1 & 5 \\
{\rm A~3125}  & 0.068 & 0.052 & 0.042 & 18.93 & 80 & 29(5) & 4 & 5 & 15 \\
{\rm A~1069}  & 0.175 & 0.134 & 0.109 & 19.07 & 56 & 18(6) & 3 & 3 & 15 \\
{\rm A~2670}  & 0.187 & 0.144 & 0.117 & 19.49 & 36 & 27 & 1 & 2 & 1 \\

            \noalign{\smallskip}\hline\noalign{\medskip}
\multicolumn{10}{l}{\rm ^{(1)}~galactic~extinction~(mag)~in~the~B,~V,~and~{\it r}~bands} \\
\multicolumn{10}{l}{\rm ^{(2)}~limiting~{\it r}~magnitudes~for~inclusion~in~the~sample} \\
\multicolumn{10}{l}{\rm ^{(3)}~number~of~early-type~galaxies~before~membership~control} \\
\multicolumn{10}{l}{\rm ^{(4)}~number~of~available~redshifts~(our~measurements~in~parenthesis)} \\
\multicolumn{10}{l}{\rm ^{(5)}~number~of~galaxies~which~are~not~cluster~members} \\
\multicolumn{10}{l}{\rm ^{(6)}~number~of~galaxies~redder~than~the~cutoff~lines~in~Fig.\ref{CM}} \\
\multicolumn{10}{l}{\rm ^{(7)}~expected~number~of~background~and~field~galaxies~} \\
         \end{array}
      \]
   \end{table*}


As explained in the previous section, the observations relative to the
present sample of nearby clusters have been done in different
observing runs, using different telescopes and different CCD cameras
(see Table~\ref{ObsLog}).  In most cases it was not possible to secure
a systematic coverage of the cluster area. In particular, for clusters
observed only during the runs~\#2 and \#3 (A2151 and A1983; detector
field of view $\sim 3^\prime\times 3^\prime$), only a few galaxies per
frame were registered.

The size and the location of the fields inside the cluster areas are
shown in Figures~\ref{Clu}a,..,g. In these figures each field color
refers to a given telescope+camera equipment (see caption) and the
grid sizes in right ascension and declination are of $30^s$ and
$5^\prime$, respectively. In Table~\ref{Fields} the list of the imaged
fields for each cluster is reported in ascending order of declination
(see pointing coordinates in columns~4 and 5) and each field is
identified by a letter (column~2). In the same table the seeing
(column~6) and the relative uncertainty of the background (column~7;
see Section 4.1) are reported.

\subsection{Selection of galaxies}

Even if the available imaging did not allow us to deal with complete
samples of galaxies as far as the cluster coverage is concerned, we
decided to set the absolute magnitude limit $M_r^{lim}=-18$~mag for
inclusion in the final sample, in order to provide homogeneous data to
study the Kormendy relation in a consistent way. That luminosity limit
represents a compromise between depth of the sampling and the
possibility to perform a detailed morphological analysis. The
corresponding limits in apparent magnitude, $r_{lim}$, given our
choice of cosmology (see Section~1) and taking into account the proper
galactic extinctions (see Table~\ref{NearClu}), are given in column 5
of Table~\ref{NearClu}.

The automatic tool SExtractor (Bertin and Arnouts~1996) was used to
produce preliminary galaxy catalogs from the images in the $r$ band,
allowing also an easy identification and rejection of stars. The
preliminary catalog contained galaxies of all morphological types,
down to SExtractor magnitudes $r_{lim} + 0.5$~mag. The additive factor
0.5~mag represents an upper limit of the bias affecting SExtractor
magnitudes of early-type galaxies (Fasano and Filippi~1998,
Franceschini et al.~1998). It prevented faint galaxies with lower
average surface brightness from being excluded {\it a priori} from the
sample.

The images were then processed by the automatic surface photometry
tool $GASPHOT$ (Pignatelli and Fasano~1999) to produce a first version
(rough but fast) of the luminosity and geometrical profiles of the
selected galaxies. These were used as a powerful complement to the
visual inspection with the IRAF--{\it imexamine} tool in estimating
the morphological types, allowing us to retain in the catalog only
galaxies classified as E or S0. Our classification scheme is not based
on quantitative morphology. However it is worth noticing that, relying
on the same observational material presented here, we give in
Fasano~et~al.~(2000) the morphological type of galaxies in nine
clusters with 0.1$\leq$z$\leq$0.25. In that paper it is shown that our
classification scheme turns out to be a robust one, both in an
absolute sense and relative to the scheme by Dressler~et~al.~(1997).
In column~6 of Table~\ref{NearClu} the number of galaxies after this
preliminary selection is reported for each cluster. Moreover, in
column~3 of Table~\ref{GalPhot} we give the morphological type of the
galaxies in the final sample.

\subsection{Cluster membership}

The next step was to evaluate the cluster membership of the galaxies
in our catalogues. The definitive criterion is indeed the redshift, so
we searched the literature for the redshift information relative to
our low redshift clusters.  To the collected 147 redshifts we have to
add the 26 new redshifts that were obtained in the framework of our
long--term project aimed at measuring line strengths of galaxies in
nearby clusters (136 low resolution spectra of galaxies in 11 nearby
clusters; Moles et al.~2001, in preparation). In column~7 of
Table~\ref{NearClu} we report, for each cluster, the total number of
available redshifts, while the number of redshifts derived from our
spectra are reported in parenthesis. The number of galaxies which are
not cluster members ($\vert cz-<cz>\vert\ge 2500$) in each cluster is
reported in column~8 of Table~\ref{NearClu}.


\begin{figure*}
    \vspace{0cm}
    \hbox{\hspace{0cm}\psfig{figure=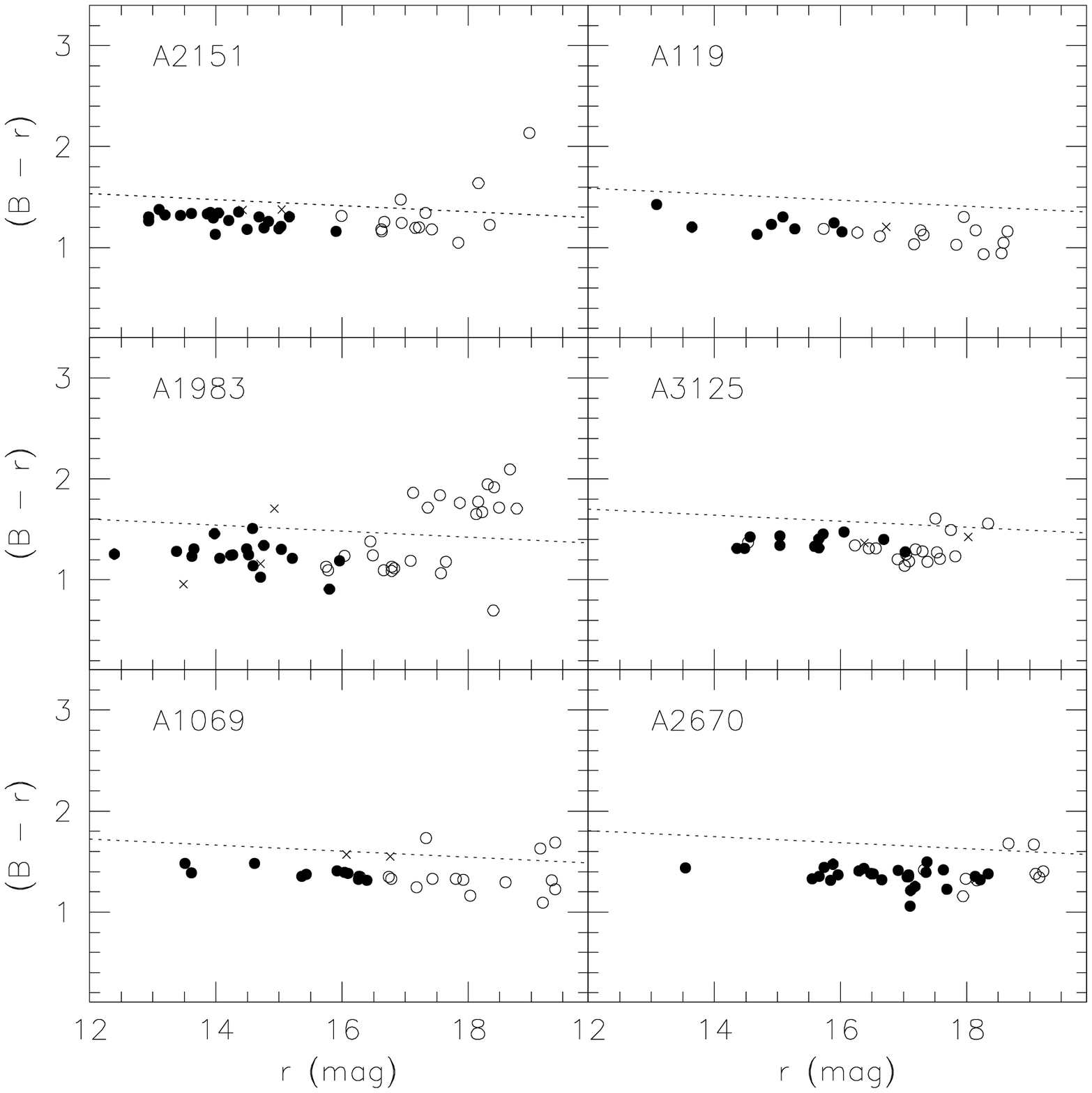,width=8.5cm}\hspace{0cm}
    \psfig{figure=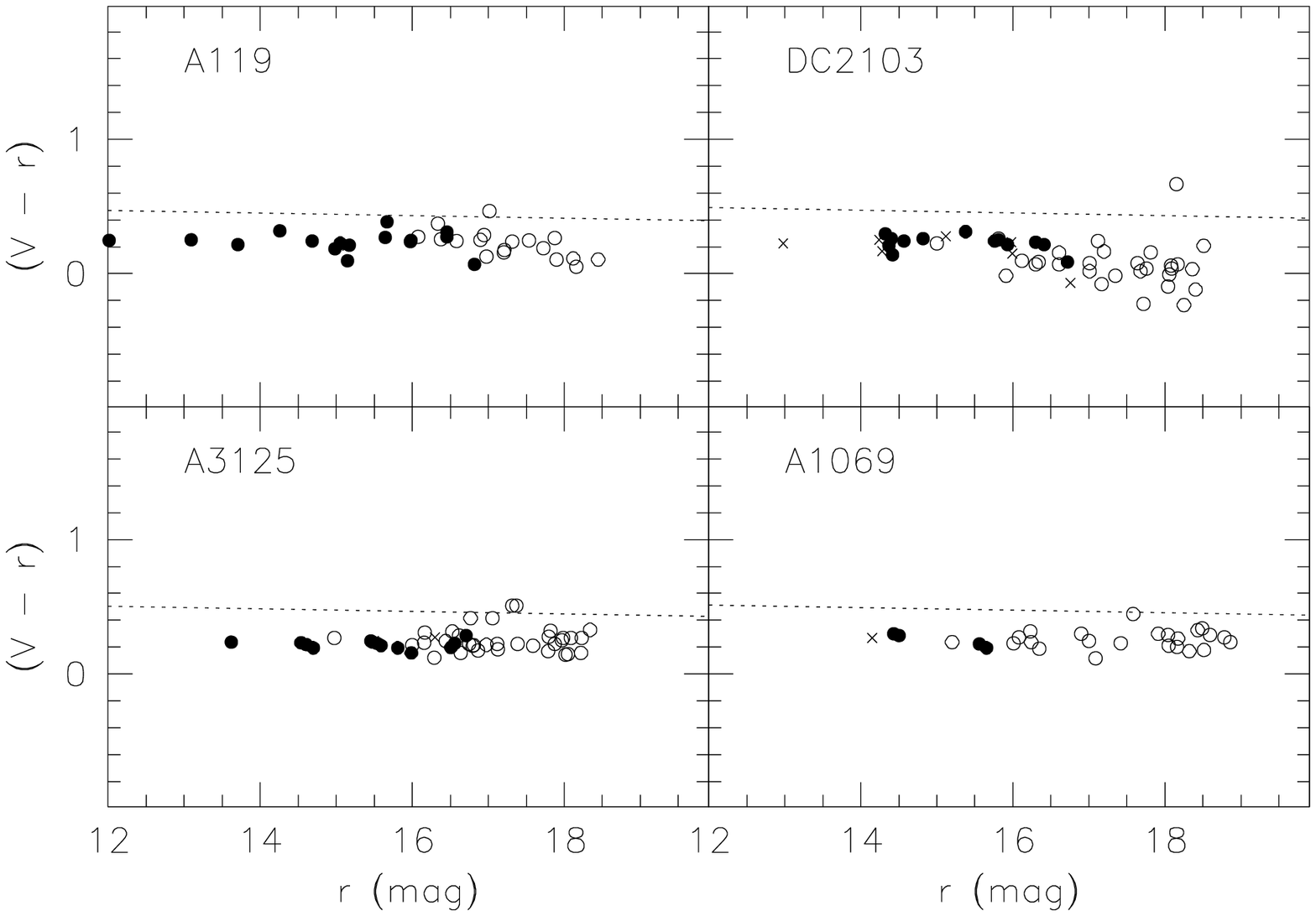,width=8.5cm}}
    \vspace{0cm}
    \caption{
Color-magnitude relations $(B-r)-r$ and $(V-r)-r$ for the nearby clusters.
The dotted lines represent the upper limits we adopted for inclusion in the
final catalog (see text). Full and open circles indicate galaxies with known
and unknown redshift, respectively. Crosses represent galaxies which, according
to the above redshift criterion, are not cluster members (see text).
    }
    \label{CM}
\end{figure*}


Since redshift determinations are usually available only for $r\ls
16$, we tried the color--magnitude relations ($CMR$) to eliminate some
red objects, presumably background galaxies. The package SExtractor
was again used to derive the instrumental colors using apertures
corresponding to a rest-frame radius of $\sim$5~kpc. Then, the
standard ($B-r$) and/or ($V-r$) colors were derived, depending on the
bands in which each cluster (or even each field) had been imaged.



In Figure~\ref{CM} we present the $CMR$s of the 7 clusters, together
with the red cutoff lines we used for the membership acceptance. These
were obtained shifting the average $CMR$s we derived from the
literature by an arbitrary factor (0.2) accounting for both the
intrinsic scatter of the $CMR$s ($\sim$0.1) and the expected
uncertainties in the magnitude estimates of SExtractor ($\sim$0.1 in
$r$ and $\sim$0.2 in $B$ and $V$). The average $CMR$s in the left part
of Figure~\ref{CM} ($B-r$~vs.~$r$) were derived adopting for the slope
the fixed value of -0.03, obtained from the data in J\o rgensen et
al.~(1995; see also Gladders et al.~1998). The zero points as a
function of redshift were computed according to the equation given in
Yee et al.~(1999) and using the transformations to the Gunn system
provided by J\o rgensen (1994). The same transformations were used to
derive the slope and the zero points for the ($V-r$)--$r$ relations
(right part of the figure). All the galaxies in each cluster redder
than the corresponding cutoff line were excluded from the final
sample; they are reported in column~9 of Table~\ref{NearClu}. The
full/open dots in the figures represent galaxies with/without measured
redshift, while the crosses indicate galaxies which, according to the
redshift criterion, are not cluster members. Notice that we do not use
any cut-off in the blue side to avoid arbitrarily eliminating genuine blue
galaxies.
We stress that we don't use the CMR as real membership discriminator,
but only as a tool to eliminate those galaxies most likely in the
background of the cluster.

In column 10 of Table~\ref{NearClu}, for each cluster (with the proper
area coverage and limiting magnitude) we report the expected
background and field contamination obtained using the galaxy number
counts given by Metcalfe~et~al.~(1995) and assuming the canonical
breakdown into morphological classes given for the field by
Dressler~(1980; E:S0:Sp+Irr~=~10:10:80). Table~\ref{NearClu} (see
also Figure~\ref{CM}) shows that, in all clusters but Abell~1983, the
expected number of background and field galaxies agrees, within the
Poissonian uncertainty, with the total number of objects excluded from
the sample due to redshift and/or color discrepancy. The sizeable
number of faint, red galaxies in Abell~1983 could indicate the
presence of some background galaxy concentration.

In the Figures~\ref{Clu}a,..,g the selected galaxies are marked by
small circles, whereas the corresponding Figures~\ref{Cluf}a,..,g show
the detailed maps of the fields, the galaxies being numbered in
ascending order of declination for each field. In this way each galaxy
in our sample is identified by the cluster name, the field letter and
the galaxy number. In Tables~\ref{GalPhot}a,..,g we report, for each
cluster, the galaxy sample for which accurate surface photometry has
been achieved.

\section{Surface photometry}

Detailed surface photometry was
obtained using the $AIAP$ package running at the Padova Observatory
(Fasano 1990), which allows to derive photometric and structural
profiles of individual galaxies. The advantages of using this software
have been presented elsewhere (i.e. Fasano
\et ~1996). We note that, due to its high degree of interactivity, the
$AIAP$ package turns out to be particularly useful for analyzing the isophotes
of galaxies embedded in high density regions, such as rich clusters or compact
galaxy groups. 

\subsection{Ellipse fitting profiles}

The surface photometry was always accomplished on the $r$ images,
apart from that relative to the observations of A119 during the runs
\#1 and \#5 (33 galaxies), for which the images in the V band turned
out to be deeper.

Before starting the analysis of individual galaxies, each frame was
handled in order to achieve a careful sampling, fitting and removal of
the sky. Apart from a few cases, a two--dimensional, first--degree
polynomial was sufficient to give an accurate fit to the sky. The
distribution of the residuals in the frame was used to estimate the
relative uncertainty of the sky level ($\Delta$sky/sky; see column~7
in Table~\ref{Fields}), which has influence on the errors of the
profiles and on the global parameters.

In each $AIAP$ run we sliced the galaxy image with a fixed step in
surface brightness (0.2~mag), we fitted the isophotes with
ellipses and we produced a set of profiles (surface brightness $\mu$,
coordinates of the center, ellipticity $\varepsilon$, position angle
$\theta$ and coefficients of the Fourier analysis of the residuals) as
a function of the semi-major axis $a_{maj}$ of the ellipses. Following
Fasano and Bonoli (1990), the error estimates of luminosity,
ellipticity and position angles profiles were derived taking into account
the $FWHM$ and the above mentioned uncertainty of the sky
level.


   \begin{figure}
      \hspace{-1cm}
      \resizebox{10cm}{!}{\includegraphics{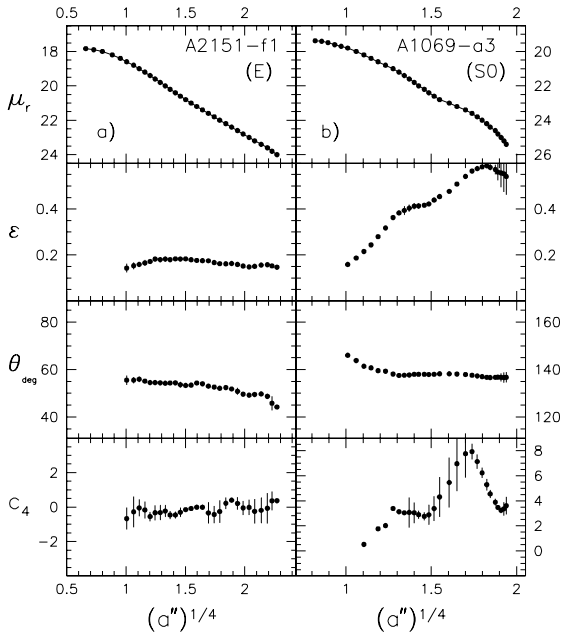}}
      \vspace{-1cm}
      \caption{
Examples of luminosity and morphological profiles extracted using
$AIAP$: $a)$ a typical elliptical galaxy (A2151-f1) and $b)$ a typical
S0 galaxy (A1069-a3).
      }
      \label{Aiap}
    \end{figure}


The errors of the Fourier coefficients were estimated from the local
noise of the profiles. To illustrate what kind of information we are
dealing with, we report in Figure~\ref{Aiap} two examples of $AIAP$
profiles, the first one referred to an elliptical galaxy (A2151--f1),
the other one to an S0 galaxy (A1069--a3). The complete set of
profiles (Tables and Figures) for the whole galaxy sample is available
in electronic form. We note that, for historical reasons, the position
angles in the Tables (see also Tables~\ref{GalPhot}a,..,g) are counted
clockwise from the North, while in the figures (see also Figure~\ref{Aiap})
they are counted counter-clockwise (again from the North).

\subsection{Provisional magnitudes and metric sizes}

The global parameters of galaxies have been mostly extracted from the
equivalent luminosity profiles, which give the surface brightness as a
function of the isophotal equivalent radius $R_{eq} = a_{maj}
\sqrt{1-\varepsilon}$.

Estimating the total magnitude $m_{_T}$ of elliptical galaxies is
known to be a rather difficult task, particularly in the cores of rich
clusters. The light distribution of elliptical galaxies smoothly
decreases outwards and it is practically impossible to establish the
galaxy bounds. The problem is two-fold: first, an extended aperture
photometry turns out to be unfeasible in the crowded galaxy fields
typical of the cluster cores; and, on the other hand, due to the
rapidly decreasing S/N ratio, usually the luminosity profiles obtained
from ellipse fitting of the isophotes cannot be extended to approach
the true value of the total magnitude close enough. This makes some
extrapolation unavoidable.

Our strategy in estimating the provisional values of the total galaxy
magnitudes was the following: ($a$) for $R_{eq}>3\times FWHM$, the
luminosity profiles were tentatively fitted with a generalized
de~Vaucouleurs law $\mu(R)=\mu_0+C(n)\times(R/R_e)^{1/n}$ (Sersic
1968, Ciotti 1991, Caon \et 1993), providing a first guess of the
parameter $n$.
To each profile fitting we assigned arbitrarily a quality index Q 
(good fit: Q=1) that takes into
account the problems encountered during the fit (such as the presence
of undulations, bumps, etc.) and during the data reduction (the galaxy
was in a crowded region, or near the boundaries of the chip,
etc.); ($b1$) in case of good fit (Q=1), the
$R^{1/n}$ law was used to extrapolate the luminosity profile in order
to derive the total magnitude; ($b2$) if the fit was not satisfactory
in spite of the large angular size and brightness of the galaxy
(luminosity profile intrinsically not performable by a $R^{1/n}$ law,
i.e. S0s with bulge+disk profiles, bright extended halos, etc..) the
extrapolation was achieved using the de~Vaucouleurs $R^{1/4}$ law or
the exponential law, depending on the shape of the outer part of the
profile; ($b3$) if the fit was not good due to the small size and/or
low brightness of the galaxy, the total magnitude was computed by
averaging the luminosities derived by the $R^{1/4}$ and exponential
extrapolations of the luminosity profile.

In order to give model independent estimates of the scale radius of
galaxies, as well as of the average surface brightness inside that
radius (both to be used in the scaling relations), we decided not to
use the effective radius $R_e$ defined by the slope of the $R^{1/n}$
law representation of the luminosity profile. Instead, we preferred to
use the half-light radius $R_{50}$, defined as the equivalent radius
enclosing half of the total galaxy light, and the corresponding
average surface brightness $\50mu$. We also derived from the
equivalent luminosity profile of each galaxy the radius $R_{75}$
corresponding to 75\% of its total luminosity (together with the
corresponding average surface brightness $\75mu$) and the Petrosian
(1976) radii $R_{P139}$ and $R_{P200}$, corresponding to the radii for
which the difference $\mu-\avmu$ (local minus average surface
brightness) is 1.39 and 2.00, respectively. We refer the reader to KJM
for an exhaustive discussion of the features of the Petrosian metric
sizes. Here we remind that, in case of a perfect de~Vaucouleurs
luminosity profile, the Petrosian radii $R_{P139}$ and $R_{P200}$
coincide with $R_{50}$ and $R_{75}$, respectively.

\subsection{Luminosity profile restoration}

The study of the Kormendy relation of early type galaxies in clusters
requires accurate estimates of some metric radius $R_m$ and of the
average surface brightness inside that radius $\mmu$. In Fasano \et
(1997) it is illustrated, for two clusters at z$\sim$0.25 (A2111 and
A1878), the crucial role played by the deconvolution of the luminosity
profiles in recovering the global parameters $R_{50}$ and $\50mu$ from
ground--based material (even of excellent quality). The redshift of
the clusters in the present sample is smaller, but the correction is
still important for the faint (and small) end of the galaxy
population, for which the values of $R_m$ and $\mmu$ derived directly
from the observed luminosity profiles can be strongly affected by the
seeing.


   \begin{figure*}
      \vspace{-5cm}
      \hspace{1cm}
      \resizebox{14cm}{!}{\includegraphics{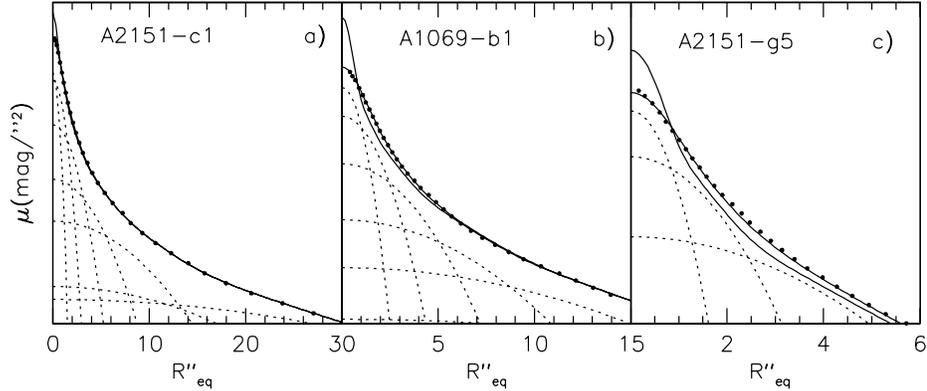}}
      \vspace{-3cm}
      \caption{
Examples of $EMGDEC$ deconvolution: $a)$ a large galaxy (A2151-c1),
$b)$ a medium size galaxy (A1069-b1) and $c)$ a small galaxy
(A2151-g5). The full dots represent the luminosity profiles extracted
by $AIAP$ and interpolated by means of suitable multi--gaussian
expansions (dotted lines). The steeper (full) lines represent the
deconvolved luminosity profiles. The surface brightness is in
arbitrary units.
      }
      \label{Decon}
    \end{figure*}


As in Fasano \et (1997), we used the Multi--Gaussian Expansion
deconvolution technique ($EMGDEC$, Bendinelli 1991) to achieve the
luminosity profile restoration. The input data of the $EMGDEC$
algorithm are the equivalent luminosity profiles of the galaxy and of
the $PSF$, both represented in analytical form by means of suitable
series of gaussians.  Although using a parametric approach, $EMGDEC$
has the advantage of a very accurate representation of the luminosity
profiles. 
In Figure~\ref{Decon} we show some examples of $EMGDEC$ restoration
applied to the luminosity profiles of galaxies belonging to our sample
and spanning a wide range in size. Obviously for large
galaxies, only the inner part of the profile is modified by $EMGDEC$,
while for small galaxies the restoration affects the whole
profiles. We emphasize that in any case the convolution of the
restored luminosity profiles perfectly reproduces the profiles
actually observed.

\begin{figure}
      \vspace{-1.5cm}
      \resizebox{9cm}{!}{\includegraphics{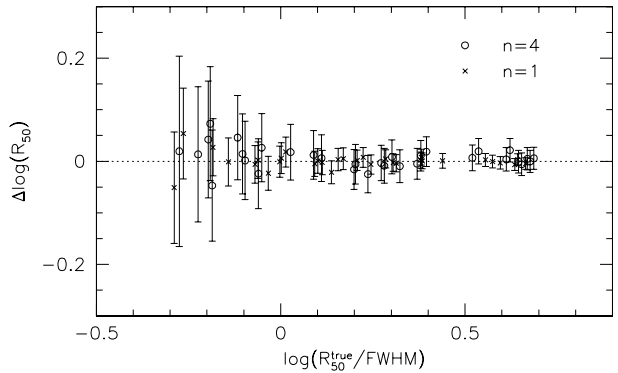}}
      \vspace{-1.5cm}
      \caption{
Difference between estimated and true value of the average half-light
radius as a function of the true value itself for a sample of 64 toy
galaxies (see text for details on the simulation). The open circles
refer to toy galaxies with de~Vaucouleurs luminosity profiles, while
crosses refer to exponential luminosity profile galaxies.
      }
      \label{DeconCheck}
    \end{figure}

It is worth noticing that the $EMGDEC$ deconvolution is not unique,
the result depending on the multi--gaussian representation (and
extrapolation) of the profiles, as well as on the so called
regularization tool (see Bendinelli 1991 for details). In order to
check the reliability and the robustness of the deconvolved half--light
radii we have analysed a sample of 64 toy galaxies with ellipticity
$\varepsilon$=0, Sersic's indexes $n$=1 and $n$=4 and half--light
radii spanning the range $R_{50}^{true}/FWHM$=0.5--5. The simulated
frames reproduces the typical conditions of our observing runs \#2 and
\#3, including the background noise. The luminosity profile of each
toy galaxy was deconvolved nine times changing both the number of
gaussians used to represent it (3 to 5) and the regularization
coefficient of $EMGDEC$ (0.0001, 0.001 and 0.01).  In
Figure~\ref{DeconCheck} we report, for each toy galaxy, the average
value (together with the relative $r.m.s.$) of the difference between
estimated and true values of the half--light radius as a function of
the true value itself. We conclude that, at least for circularly symmetric
objects, the $EMGDEC$ tool produces
un--biased values of the deconvolved half--light radii and that,
changing the starting conditions within wide ranges produces only
marginal changes in the half--light radius down to $R_{50}\simeq
FWHM$. 


\begin{figure}
      \hspace{-2.5cm}
      \resizebox{14cm}{!}{\includegraphics{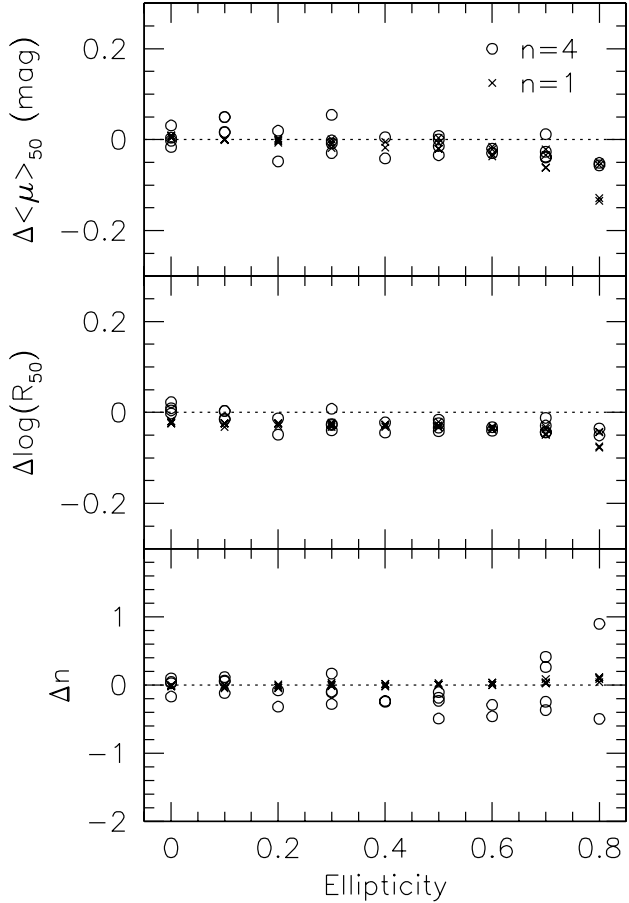}}
      \caption{
Difference between estimated and true values of the average half-light
surface brightness, half-light radius and Sersic's index, after $EMGDEC$
restoration, for a sample of 64 toy galaxies (see text for details on the
simulation). The open circles refer to toy galaxies with de~Vaucouleurs
luminosity profiles, while crosses refer to exponential luminosity profile 
galaxies. 
      }
      \label{Elli}
    \end{figure}

Since the $EMGDEC$ algorithm only deals with circularly symmetric
objects, we have carried out numerical simulations of galaxies with
different flattening in order to explore how the ellipticity
influences the $EMGDEC$ equivalent profile restoration, in particular
as far as the Sersic's index, the effective radius and the
corresponding average surface brightness estimates are concerned.
Figure~\ref{Elli} illustrates the results for a sample of 64
elliptical ($n$=4) or disk ($n$=1) toy galaxies having faint magnitudes
(17.5-18.5 mag) and small effective radii (5-9 pixels). Again, the simulated
frames reproduces the typical conditions of our observing runs \#2 and
\#3. We conclude that the influence of the isophotal flattening on the
estimation of the equivalent parameters after $EMGDEC$ restoration is
relatively unimportant except for an ellipticity greater than 0.7.


\section{Global parameters and errors}

We have used the restored, $equivalent$ luminosity profiles to derive
the final values of the global parameters related to magnitudes and
metric sizes, as well as to compute the final value of the Sersic's
index $n$. The other '$globals$' related to the morphology
($\varepsilon$, $\theta$ and Fourier coefficients) have been derived
from the original profiles, since our one-dimension technique of
restoration is not able to recover the '$true$' galaxy geometry.

\subsection{Final magnitudes and metric sizes}

The same strategy as outlined in Section~4.2 to estimate the
$provisional$ total magnitudes from the raw luminosity profiles, was
used to estimate the final total magnitudes from the de-convolved
luminosity profiles (columns 7 of Tables~\ref{GalPhot}). In this case
we have adopted a fixed range of $R/R_{50}$ (0.15 to 4, with a minimum
allowed angular radius of 0$\asec$5) to produce $\chi^2$ fits of the
luminosity profiles, leading to the final values of the Sersic's
parameter $n$. Since the value of $R_{50}$ depends on the total
magnitude itself, a two--step iterative procedure was necessary. We
note that, if the quality of the fit is good ($Q=1$, case [$b1$]), the
total magnitude $m_{_T}$ was derived extrapolating the observed
luminosity profile by a generalized de~Vaucouleurs law with index
$n$. Otherwise (cases [$b2$] and [$b3$] in Section~4.2) the $R^{1/4}$
and/or the exponential laws were used for the extrapolation.

The final values of the different metric sizes mentioned in Section
4.2, together with the corresponding average surface brightness, were
easily derived from the restored luminosity profiles, once the total
magnitudes were known.
It is worth mentioning that the restored values of the total magnitude
differ only slightly from the corresponding ones provisionally derived
from the raw luminosity profiles. On the contrary, the restoration
procedure deeply influences the different kinds of metric radius, as
well as the corresponding average surface brightness.

\begin{figure}
      \vspace{-4cm}
      \hspace{-2.5cm}
      \resizebox{14cm}{!}{\includegraphics{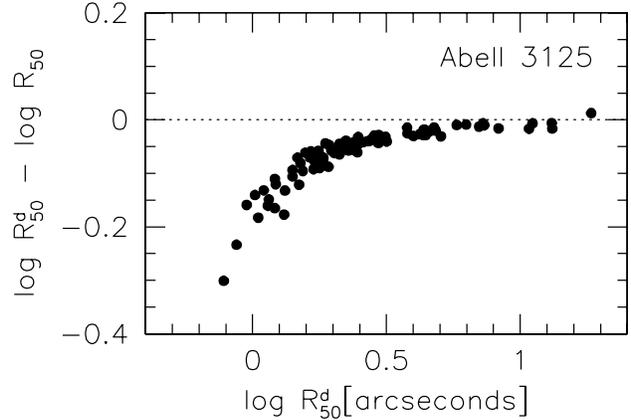}}
      \vspace{-3cm}
      \caption{
Difference between half--light radii before and after
deconvolution of the luminosity profiles for the galaxies in Abell~3125.
      }
      \label{DeconRad}
    \end{figure}

As an example, in Figure~\ref{DeconRad} the
difference between restored and raw values of $R_{50}$ for the galaxies in
A3125 is reported as a function of $R_{50}$. The same qualitative behaviour is
found for all clusters, with systematic differences mainly depending on the
seeing, on the pixel-size and on the S/N ratio, and individual differences due
to the shape of profiles.


In order to give realistic estimates of the uncertainties associated
to total magnitude, metric radius and average surface brightness, we
have taken into account all the possible sources of error. The
contribution due to the standard calibration includes all the
uncertainties related to the color term and to the atmospheric
extinction and can be easily derived from Table~\ref{CalCoef}
(column~7). Even affecting all the quantities related to some
magnitude estimate (i.e. surface brightness), this contribution does
not influence the shape of the luminosity profiles and, therefore, it
does not affect any estimate of the metric radius. On the contrary,
the uncertainty related to the shape of the luminosity profile
(including that attributable to the background removal) and to its
extrapolation, affects both the magnitude and the radius. In case of
luminosity profiles well represented by a generalized de~Vaucouleurs
law (case [$b1$] in Section~4.2), we estimated this contribution by
examining the $r.m.s$ of the $R^{1/n}$ fit as a function of $n$. The
behaviour of this function varies with the profile. In particular,
when $n$ is large ($n>$4), the minimum of the function $rms(n)$ is
rather flat, while for small values of $n$ it turns out to be much
better defined. In any case, the best fit $r.m.s$ turns out to span
the range 0.02$-$0.04. We empirically verified that an excess of $\Delta_{rms}$=0.01~mag
with respect to the minimum, always corresponds to a value of $\Delta
n$ which is large enough to make the new fit with $n^\prime =
n\pm\Delta n$ significantly worse than the best fit. We therefore fixed, for each galaxy,
the confidence range $\Delta n$ corresponding to $\Delta_{rms}$=0.01,
and we derived the expected uncertainties for the total magnitude
($\Delta m_{_T}$), metric size ($\Delta R$) and surface brightness
($\Delta\mu$). In case of a bad fit of the luminosity profile (cases
[$b2$] and [$b3$] of Section~4.2), we simply assumed the magnitude
difference between the $R^{1/4}$ and exponential extrapolations to be
an estimate of $\Delta m_{_T}$, and derived the corresponding values
of $\Delta R$ and $\Delta\mu$. 

\begin{figure}
      \hspace{-1cm}
      \resizebox{9cm}{!}{\includegraphics{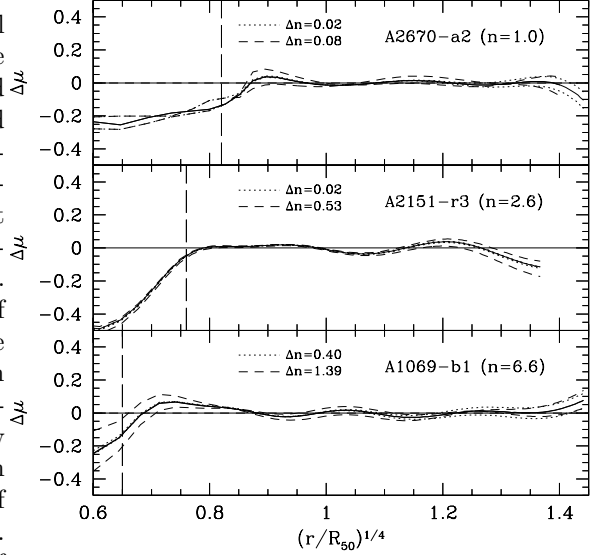}}
      \caption{
Surface brightness residuals of best fitting Sersic's law for three
galaxies with very different $n$. The meaning of the dashed and dotted
lines is explained in the text.
      }
      \label{SersFit}
    \end{figure}

We illustrate in Figure~\ref{SersFit} the adequacy of our fits to the
luminosity profiles. We present the results for 3 galaxies with very
different Sersic indices. The fraction of the profile actually used
for the fits is indicated with the vertical, dashed lines in each
panel.  The best fit, obtained with the prescriptions given above, is
given by the full lines. The acompanying dashed lines are for the fits
corresponding to $n^\prime = n\pm\Delta n$, n being the Sersic
coefficient of the best fit. Finally, the dotted lines show the effect
of the errors in the background (given in Table~\ref{Fields}).



\subsection{Ellipticity, isophotal twisting and Fourier coefficients}

Apart from the global quantities extracted from the equivalent
luminosity profile ($n$, $m_{_T}$, $R_{50}$, $\50mu$, etc.), we
produced some more '$globals$' related to the geometrical profiles. In
particular, we measured the ellipticity, the position angle and the
Fourier coefficient $c_4$ (disky/boxy) of the equivalent effective
isophote and at $R_{eq}=3\times R_{50}$.  We also recorded the maximum
ellipticity found on the profile ($\varepsilon_{max}$), the most
likely value of the isophotal twisting ($\Delta\theta$), together with
the minimum ($\Delta\theta_{min}$) and maximum ($\Delta\theta_{max}$)
values allowed on the basis of the position angle uncertainties, and a
luminosity--weighted value of $c_4$ ($c_4^w$), together with the
minimum and maximum values of $c_4$ found on the profile ($c_4^{min}$,
$c_4^{max}$). Finally, again from the luminosity profiles, we derived
three different gradients of surface brightness, namely:

$\delta\mu_1=(\mu_{R_{50}}-\mu_{R_{50}/2})$,

$\delta\mu_2=(\mu_{R_{50}/2}-\mu_{R_{50}/3})$,

$\delta\mu_3=(\mu_{R_{50}/4}-\mu_{R_{50}/3})$.

\medskip\noindent
All the quantities relevant for our analysis are reported for each galaxy
in Tables~\ref{GalPhot}a,..,g. The electronic versions of these tables
contain the whole information, whereas their printed versions contain
only the most important entries, namely:

\noindent{\it Column 1}: galaxy identifier;

\noindent{\it Column 2}: galaxy name [NGC/IC/UGC/Dressler(1980)];

\noindent{\it Column 3}: morphological type;

\noindent{\it Column 4}: right ascension (J2000);

\noindent{\it Column 5}: declination (J2000);

\noindent{\it Column 6}: redshift (an asterisk indicates our own measurement);

\noindent{\it Column 7}: (m$_T$) total magnitude in the Gunn~r band;

\noindent{\it Column 8}: ($\Delta$m$_T$) 1$\sigma$ uncertainty on the total magnitude;

\noindent{\it Column 9}: color $(B - r)$;

\noindent{\it Column 10}: color $(V - r)$;

\noindent{\it Column 11}: (R$_{50}$) equivalent half--light radius in arcseconds, after deconvolution;

\noindent{\it Column 12}: ($\Delta$R$_{50}$) 1$\sigma$ uncertainty on the deconvolved equivalent half--light radius;

\noindent{\it Column 13}: ($\50mu$) mean surface brightness relative to the deconvolved half--light radius;

\noindent{\it Column 14}: ($\Delta\mu_{50}$) 1$\sigma$ uncertainty on the mean surface brightness relative to the deconvolved half--light radius;

\noindent{\it Column 15}: ($n$) Sersic's index of the deconvolved luminosity profile;

\noindent{\it Column 16}: ($\Delta n$) 1$\sigma$ uncertainty on the Sersic's index;

\noindent{\it Column 17}: (Q) quality index of the fit (good fit = 1);

\noindent{\it Column 18}: ($\varepsilon_e$) ellipticity of the half--light isophote;

\noindent{\it Column 19}: ($\varepsilon_{max}$) maximum ellipticity;

\noindent{\it Column 20}: ($\theta_e$) position angle (clockwise from the North) of the half--light isophote; 

\noindent{\it Column 21}: ($\Delta\theta$) average isophotal twisting estimated from the position angle profile;

\noindent{\it Column 22}: [$c_4(r_e)$] disky/boxy Fourier coefficient ($\times$100) of the half--light isophote;

\noindent{\it Column 23}: ($c_4^w$) luminosity weighted value of the Fourier coefficient;

\noindent{\it Column 24}: ($\delta\mu_1$) surface brightness gradient ($\mu_{R_{50}}-\mu_{R_{50}/2}$);

\noindent{\it Column 25}: ($\delta\mu_2$) surface brightness gradient ($\mu_{R_{50}/2}-\mu_{R_{50}/3}$).

\medskip
All the photometric and morphological parameters will be used in a
forthcoming paper of the series to investigate the possibility to
define some photometric version of the fundamental plane of
early--type galaxies.

\section{Internal and external comparisons}

\subsection{Check for consistency}

To check the robustness of our results, we compared the surface
photometry from different runs. First, we compared the different
magnitudes for each standard galaxy. Table~\ref{SGSam} lists the
sample of standard galaxies we observed. For each galaxy we report the
runs when it was observed, together with the average values we found
for the total $r$ magnitude, half--light radius and Sersic's index
$n$. 


   \begin{table*}
      \caption[]{The sample of standard galaxies}
         \label{SGSam}
      \[
         \begin{array}{llllllllcrcc}
            \hline\noalign{\smallskip}

\multicolumn{1}{c}{\rm Name}&
\multicolumn{1}{c}{\rm Type}&
\multicolumn{3}{c}{\alpha (2000)}&
\multicolumn{3}{c}{\delta (2000)}&
\multicolumn{1}{c}{\rm runs}&
\multicolumn{1}{c}{r_{_T}^{(1)}}&
\multicolumn{1}{c}{R_{50}^{(2)\prime\prime}}&
\multicolumn{1}{c}{n^{(3)}}\\
\noalign{\smallskip}\hline\noalign{\smallskip}
{\rm NGC~1199} & E3 & 03^{h} & 03^{m} & 38.^{s}6 & -15^{\circ} &
36^{\prime} & 51^{\prime\prime} & 4 & 11.38 & 25.5 & 5.5\pm 0.07 \\
{\rm NGC~1395} & E2 & 03 & 38 &  29.6 & -23 & 01 & 40 & 1,4,5 & 9.89 & 48.1 & 5.0\pm 0.04 \\
{\rm NGC~1407} & E0 & 03 & 40 &  11.8 & -18 & 34 & 48 & 1 & 9.70 & 74.4 & 6.6\pm 0.12 \\
{\rm NGC~1439} & E1 & 03 & 44 &  49.9 & -21 & 55 & 13 & 1 & 11.18 & 33.2 & 9.7\pm 0.08 \\
{\rm NGC~1726} & S0 & 04 & 59 &  41.8 & -07 & 45 & 16 & 6 & 11.19 & 29.7 & 3.2\pm 0.06 \\
{\rm NGC~2340} & E & 07 & 11 &  10.8 & +50 & 10 & 28 & 6 & 11.34 & 51.0 & 3.1\pm 0.03 \\
{\rm NGC~2986} & E2 & 09 & 44 &  16.0 & -21 & 16 & 41 & 5 & 10.51 & 41.0 & 9.5\pm 0.09 \\
{\rm NGC~4839} & S0 & 12 & 57 &  24.2 & +27 & 29 & 54 & 2 & 11.88 & 37.4 & 2.9\pm 0.11 \\
{\rm NGC~4841A} & E & 12 & 57 &  32.0 & +28 & 28 & 38 & 2,3 & 12.20 & 20.0 & 6.8\pm 0.14 \\
{\rm NGC~4841B} & E & 12 & 57 &  33.9 & +28 & 28 & 54 & 2,3 & 13.11 & 12.0 & 4.6\pm 0.04 \\
{\rm NGC~4874} & E & 12 & 59 &  35.9 & +27 & 57 & 31 & 2 & 11.07 & 77.5 & 3.5\pm 0.06 \\
{\rm NGC~4889} & E4 & 13 & 00 &  07.7 & +27 & 58 & 33 & 2,3 & 11.03 & 30.0 & 5.5\pm 0.13 \\

            \noalign{\smallskip}\hline\noalign{\medskip}
\multicolumn{12}{l}{\rm
^{(1)}~average~total~apparent~magnitude~in~the~Gunn~{\it r}~band} \\
\multicolumn{12}{l}{\rm ^{(2)}~average~half~light~radius~in~arcseconds} \\
\multicolumn{12}{l}{\rm ^{(3)}~best~fit~Sersic's~index~and~relative~uncertainty} \\
         \end{array}
      \]
   \end{table*}

\begin{figure}
      \vspace{-4.5cm}
      \hspace{-1cm}
      \resizebox{11.5cm}{!}{\includegraphics{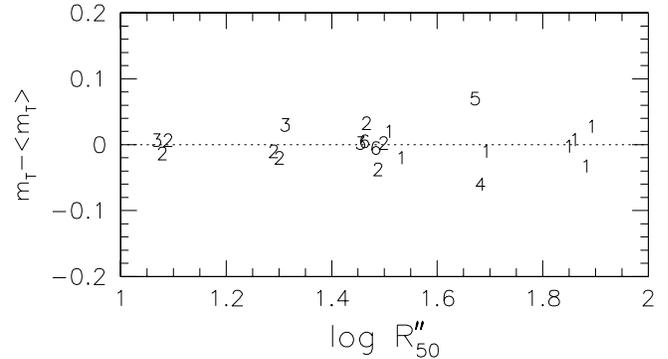}}
      \vspace{-2cm}
      \caption{
Check for internal consistency for total magnitudes of standard galaxies. The
numbers in the plot refer to the runs in which the standard galaxies have
been observed.
      }
      \label{Check0}
    \end{figure}

In Figure~\ref{Check0} we report, as a function of the half--light radius
in arcseconds, the differences between the individual magnitude measurements 
of each galaxy and the average of all available measurements for the same galaxy. 
Unfortunately, only for NGC~1395 observations in three
different runs turned out to be available, the other standard galaxies
having been observed in the same run or (at best) in two different
runs. Even if this fact prevented us from performing a complete check
for consistency of our photometric zero points, from
Figure~\ref{Check0} we concluded that the uncertainties are of the
order of a few hundredths of magnitude, at worst.


Then we compared the surface photometry of galaxies in our clusters.
The cluster Abell~119 turned out to be the best candidate for this
purpose, since it was imaged in three different observing runs
(\#1,\#5 and \#7), with three different instrumental setups (see
Tables~\ref{ClTSam} and \ref{ObsLog}).

As mentioned in Section 2.3, during run~\#1 (DFOSC~1994) we
encountered several calibration problems. The file headers were
corrupted and we were able to retrieve the calibration only for the
night Dec.~4 (see Table~\ref{CalCoef}).  In spite of these problems,
we tentatively decided to adopt the calibration coefficients relative
to this night as representative of the whole run, checking a
posteriori the possible systematic zero--point differences with the
other runs. Therefore, we retrieved some exposures in the field of
Abell~119 taken during the same run~\#1 (nights Dec.~7--8~1994) and
centered on the galaxies $D41$, $D99$ and $D105$ (our identifications:
$b$5, $e$1 and $g$4, respectively).  We also retrieved an exposure
taken during run~\#5, centered on the galaxy $D105$.

In summary, we compared imaging from the three runs~\#1, \#5 and \#7,
for the above mentioned fields. In each field more than 15 galaxies of
various sizes and luminosity turned out to be in common among the
different runs. Again the $AIAP$ tool was used to analyze the galaxies
in the retrieved frames, providing their luminosity and geometrical
profiles, as well as their global parameters.


   \begin{figure}
      \resizebox{\hsize}{!}{\includegraphics{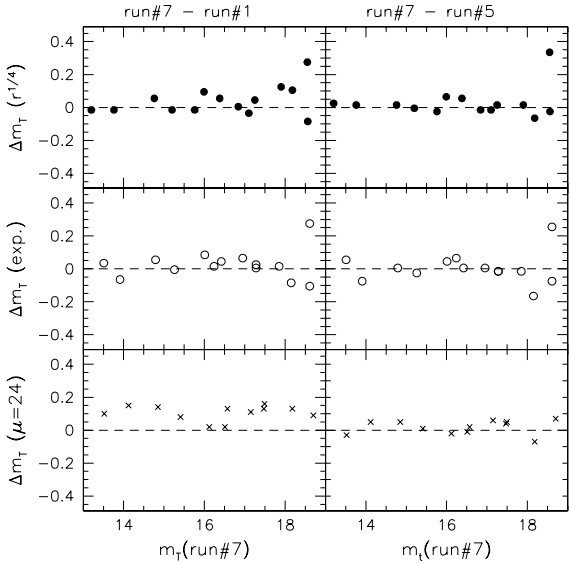}}
      \caption{
Check for internal consistency for different kinds of magnitudes.
      }
      \label{Check1}
    \end{figure}


In Figure~\ref{Check1} the differences between different kinds of
magnitudes derived for each galaxy from different runs are shown. In
particular, in the lower panels we illustrate the magnitude
differences relative to the isophote $\mu_r=24$. Being insensitive to
the adopted extrapolation of the luminosity profiles (see
Section~4.2), these differences should give indication about the true
biases affecting the different runs. As expected, the calibration
relative to run~\#1 turned out to be inaccurate. In particular,
accordingly to Figure~\ref{Check1}, a systematic zero point correction
of $\sim$0.1~mag was applied to the magnitudes of galaxies observed
during that run. In the upper and middle panels we report the
magnitude differences obtained by using the pure $R^{1/4}$ ($n$=4) and
exponential ($n$=1) extrapolations (depurated from the previous
biases), respectively. They give an indication of the maximum
uncertainties inherent to the extrapolation procedure.


   \begin{figure}
      \resizebox{\hsize}{!}{\includegraphics{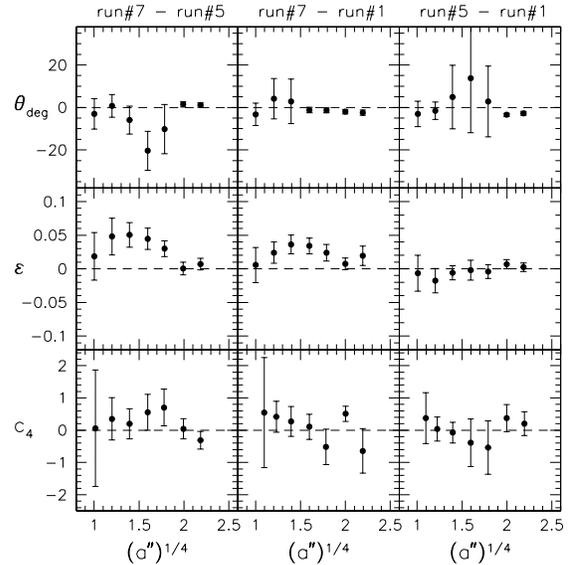}}
      \caption{
Binned and averaged differences in the $\varepsilon$, $\theta$ and $c_4$
profiles of galaxies in Abell~119 from the comparison of different observing
runs in pairs.
      }
      \label{Check2}
    \end{figure}


The binned and averaged residuals of ellipticity, position angle and
Fourier coefficient for the galaxies in common among the three runs,
are plotted in Figure~\ref{Check2} as a function of the isophotal
semi--major axis.  Figure~\ref{Check2} shows a general good agreement
among the morphological profiles of the same galaxies obtained in
different runs. The wave-like behaviour of the ellipticity residuals
in the comparisons involving run~\#7 can be easily explained by the
better seeing of this run, which is likely to produce a better
representation of the inner (possibly flat) isophotes.

\subsection{External comparisons}


   \begin{figure}
      \hspace{-1.5cm}
      \resizebox{11cm}{!}{\includegraphics{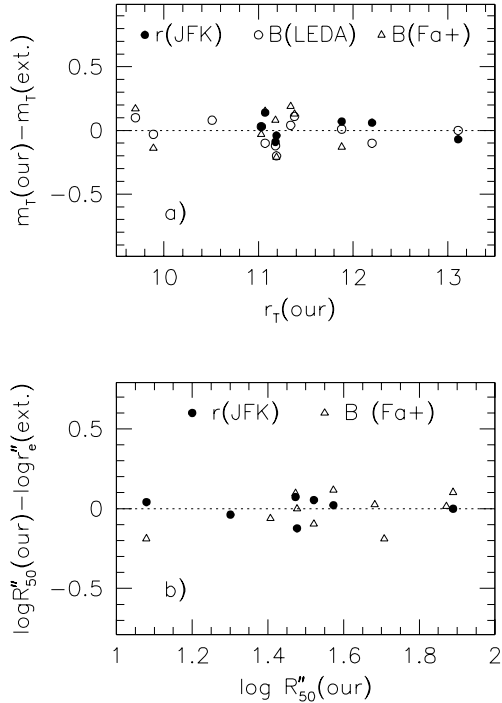}}
      \caption{
Total magnitudes and effective radii of nearby standard galaxies from
our surface photometry are compared with the values given by JF and
JFK (full dots), by Paturel et al.~(1997; open dots) and by Faber et
al.~(1987; open triangles).
      }
      \label{Ext1}
    \end{figure}


We compared our photometry with the data available in the literature
for the nearby standard galaxies we observed during the various runs
(see Table~\ref{SGSam}). In particular, in Figure~\ref{Ext1} our
photometry is compared with that of J\o rgensen and Franx (1994) and
J\o rgensen et al.~(1995; both indicated by JFK in the plots), as well
as with the data collection from Paturel et al.~(1997; LEDA) and with
the results given by Faber et al.~(1987, FA+ in the plots). Our
observations in the $r$ band were directly compared with the
corresponding ones from JFK, whereas, to compare with the magnitudes
from LEDA and FA+, we have converted them into the $r$ band by using
our colors $(B-r)$. We found:

\noindent
$<$m$_T$(our)-m$_T$(other)$>$=0.005$\pm$0.025~mag~(r.m.s.=0.112),

\noindent
$<$logR$_{50}$(our)-logR$_e$(other)$>$=0.009$\pm$0.023~(r.m.s.=0.095).

It is worth stressing that the methodologies used in the above
mentioned works to derive total magnitudes and effective radii differ
from ours and also differ from each one another. In particular, our
$R_{50}$ is defined as the equivalent radius enclosing half of the
total galaxy light (irrespectively of the shape of the luminosity
profile), while past work generally derive the effective radius $R_e$
assuming an $r^{1/4}$ profile. This produces systematically different
results, depending on the true luminosity profile shape, and 
likely explains the relatively large scatter we find in comparing
the radius estimates (22\%). Actually, an even larger scatter (30\%)
was suggested to be expected by Kelson et~al.~(2000) in their
extensive discussion of comparisons of $R_{50}$ derived from
bulge+disk fits and from Sersic's law fits.

\section{Summary and future plans}

This paper mainly deals with the illustration of the photometric data
we have collected in the framework of a long term project aimed at
investigating systematically the so called `scaling relations' of
early--type galaxies in clusters. The main goal of this project is to
throw light upon the cosmic variance of the scaling relations
themselves, as well as upon their possible dependence on the redshift.

We start here discussing the reduction and calibration procedures we
have applied to the whole photometric data set, relative to the global
sample of 22 clusters in different bins of redshift, up to
z$\sim$0.25.  Then, we present the detailed surface photometry of 312
early--type galaxies in 7 nearby clusters belonging to the first
redshift bin and defining the local reference sample. The whole set of
luminosity and geometrical profiles is placed at reader's disposal (in
both tabular and postscript format) in the electronic version of the
paper, while several global photometric and structural parameters are
evaluated for each galaxy and are collected in tables (one for each
cluster), which again are at reader's disposal.

Here we do not try to analyze the data in the tables, nor to seek for
possible correlations among the various parameters. These items will
be addressed in the third paper of this series, where the scaling
relations of early--type galaxies in the local sample of clusters will
be discussed in detail, using both the photometric and the
spectroscopic information. We will present the latter one in the
second (forthcoming) paper of this series, where the results of both
the low- and the intermediate- resolution spectroscopy for a subsample
of the present galaxy sample will be discussed.

The two final steps of the series will concern the detailed surface
photometry of early--type galaxies in the remaining 15 (more distant)
clusters and the analysis of the scaling relations as a function of
different parameters, including the redshift of the cluster.

\begin{acknowledgements}
We wish to thank the referee Dan Kelson for the very usefull suggestions
which helped us to greatly improve the final version of the paper.
\end{acknowledgements}

   \begin{table*}[p]
      \caption[]{The observed fields}
         \label{Fields}
      \[
\begin{array}{cccccccc}
            \hline\noalign{\smallskip}

  \multicolumn{1}{c}{\rm Cluster}
& \multicolumn{1}{c}{\rm Field}
& \multicolumn{1}{c}{\rm Run}
& \multicolumn{1}{c}{\alpha_c}
& \multicolumn{1}{c}{\delta_c}
& \multicolumn{1}{c}{FWHM}
& \multicolumn{1}{c}{\rm Background^{(1)}}
& \multicolumn{1}{c}{N_{gal}^{(2)}} \\
 & & & {\rm (2000)} & {\rm (2000)} & \prime\prime & {\rm rel.err.} & \\
\hline
{\rm A~2151} & $a$ & 2 & 16:02:48.8 & 17:09:59 & 0.6 & 0.005 & 2 \\
      & $b$ & 2 & 16:04:19.1 & 17:10:48 & 0.7 & 0.008 & 1 \\
      & $c$ & 3 & 16:03:31.8 & 17:11:53 & 0.7 & 0.005 & 1 \\
      & $d$ & 2 & 16:04:02.3 & 17:16:52 & 0.7 & 0.005 & 1 \\
      & $e$ & 2 & 16:05:32.6 & 17:35:54 & 0.7 & 0.005 & 1 \\
      & $f$ & 2 & 16:04:38.0 & 17:42:44 & 0.5 & 0.005 & 1 \\
\end{array}
      \]
   \end{table*}

\addtocounter{table}{-1}
   \begin{table*}[p]
      \caption[]{The observed fields (Continue..)}
         \label{Fields}
      \[
         \begin{array}{cccccccc}
            \hline\noalign{\smallskip}

  \multicolumn{1}{c}{\rm Cluster}
& \multicolumn{1}{c}{\rm Field}
& \multicolumn{1}{c}{\rm Run}
& \multicolumn{1}{c}{\alpha_c}
& \multicolumn{1}{c}{\delta_c}
& \multicolumn{1}{c}{FWHM}
& \multicolumn{1}{c}{\rm Background}
& \multicolumn{1}{c}{N_{gal}} \\
 & & & {\rm (2000)} & {\rm (2000)} & \prime\prime & {\rm rel.err.} & \\
\hline
{\rm DC~2103} & $a$ & 4 & 21:08:59.6 & -39:51:04 & 1.1 & 0.002 & 9 \\
      & $b$ & 4 & 21:06:04.5 & -39:35:42 & 1.2 & 0.003 & 17 \\
      & $c$ & 4 & 21:07:47.6 & -39:29:08 & 1.2 & 0.004 & 15 \\
\hline
{\rm A~3125} & $a$ & 5 & 03:26:43.3 & -53:50:03 & 1.2 & 0.008 & 2 \\
      & $b$ & 5 & 03:29:12.8 & -53:28:36 & 1.5 & 0.002 & 5 \\
      & $c$ & 5 & 03:27:43.5 & -53:23:34 & 1.0 & 0.003 & 16 \\
      & $d$ & 5 & 03:23:55.0 & -53:18:29 & 1.3 & 0.005 & 11 \\
      & $e$ & 5 & 03:27:01.7 & -53:08:36 & 1.6 & 0.001 & 6 \\
      & $f$ & 5 & 03:24:11.6 & -53:08:28 & 1.2 & 0.001 & 11 \\
      & $g$ & 5 & 03:23:16.4 & -53:05:28 & 1.4 & 0.002 & 5 \\
      & $h$ & 5 & 03:30:53.7 & -52:51:05 & 1.3 & 0.001 & 11 \\
      & $i$ & 5 & 03:28:02.1 & -52:40:28 & 1.4 & 0.002 & 4 \\
\hline
{\rm A~1069} & $a$ & 5 & 10:37:37.4 & -09:09:08 & 1.5 & 0.012 & 5 \\
      & $b$ & 5 & 10:40:38.3 & -08:51:27 & 1.3 & 0.004 & 2 \\
      & $c$ & 5 & 10:39:47.0 & -08:37:44 & 1.6 & 0.004 & 8 \\
      & $d$ & 5 & 10:40:05.2 & -08:22:05 & 1.2 & 0.004 & 4 \\
      & $e$ & 5 & 10:41:57.8 & -08:12:42 & 1.6 & 0.006 & 10 \\
      & $f$ & 5 & 10:38:19.3 & -08:11:42 & 1.4 & 0.004 & 12 \\
      & $g$ & 5 & 10:40:41.2 & -07:58:17 & 1.4 & 0.006 & 9 \\
\hline
{\rm A~2670} & $a$ & 7 & 23:54:16.0 & -10:25:10 & 1.1 & 0.002 & 33 \\

            \noalign{\smallskip}\hline
         \end{array}
      \]
   \end{table*}


\begin{changemargin}{-1cm}{1cm}{-1cm}{25.6cm} 
\begin{landscape}
   \begin{table}[p]
      \caption[]{{\bf a} -- Photometric and morphological parameters of galaxies in Abell~2151}
         \label{GalPhot}
      \[
         \begin{array}{rcccccccccrrcrcccccrrrrcc}

\noalign{\bigskip}\hline\noalign{\smallskip}
{\rm Galaxy} & {\rm Name} & {\rm Type} & \alpha (2000) & \delta (2000) & z & {\rm m}_{_T} & \Delta{\rm m} & (B-r) & (V-r) & {\rm R}_{50} & \Delta{\rm R}_{50} & \50mu & \Delta{\rm \mu_{50}} & n & \Delta{n} & Q& \varepsilon_e & \varepsilon_{max} & \theta_e & \Delta\theta & c_4(r_e) & {c_4^w}~~ & \delta\mu_1 & \delta\mu_2 \\
\noalign{\smallskip}
\hline\noalign{\smallskip}

 {\rm a1} & {\rm D9}  &{\rm E/S0} &16\ 02\ 49.1& 17\ 10\ 05&0.0347  &14.81&0.03&1.37&      &   4.29 &    0.20 &   19.95 &  0.06   &  2.80   &    0.14   & 0&     0.28 &     0.41 &  142.5 &  24.3 &  0.66\ \  &  0.29\ \  &     1.11 &     0.60\\
 {\rm a2} &           &{\rm S0}   &16\ 02\ 47.3& 17\ 10\ 56&        &17.07&0.02&1.31&      &   2.58 &    0.06 &   21.07 &  0.06   &  0.90 &     0.26   & 1&     0.36 &     0.48 &   16.2 &   7.1 &     1.28\ \ &     0.26\ \ &     1.14 &     0.58\\
 {\rm b1} & {\rm D4}  &{\rm E}    &16\ 04\ 19.5& 17\ 10\ 49&0.0341  &14.62&0.02&1.24&      &   4.64 &    0.18 &   19.97 &  0.06   &  3.10 &     0.47   & 1&     0.07 &     0.19 &  158.2 &  63.6 &  -0.27\ \ &   -0.23\ \ &     1.24 &     0.58\\
 {\rm c1} &{\rm N6034}&{\rm E}    &16\ 03\ 32.0& 17\ 11\ 55&0.0339  &13.23&0.12&1.44&      &  14.63 &    4.21 &   21.04 &  0.38   & 11.60 &     4.58   & 1&     0.22 &     0.26 &  120.9 &  18.2 & -0.63 \ \ &  -0.52 \ \ &     1.24 &     0.76\\
 {\rm d1} &{\rm  D15} &{\rm E/S0} &16\ 04\ 02.7& 17\ 16\ 56&0.0330  &14.49&0.02&1.42&      &   4.91 &    0.16 &   19.89 &  0.05   &  4.50 &     0.40   & 1&     0.27 &     0.38 &  136.6 &   3.4 & -0.17 \ \ &   0.06\ \ &     1.22 &     0.60\\
\end{array}
      \]
   \end{table}
\end{landscape}

\begin{landscape}
\addtocounter{table}{-1}
   \begin{table}[p]
      \caption[]{{\bf b} -- Photometric and morphological parameters of galaxies in Abell~119}
         \label{GalPhot}
      \[
         \begin{array}{rcccccccccrrcrcccccrrrrcc}

\noalign{\bigskip}\hline\noalign{\smallskip}
{\rm Galaxy} & {\rm Name} & {\rm Type} & \alpha (2000) & \delta (2000) & z & {\rm m}_{_T} & \Delta{\rm m} & (B-r) & (V-r) & {\rm R}_{50} & \Delta{\rm R}_{50} & \50mu & \Delta{\rm \mu_{50}} & n & \Delta{n} & Q& \varepsilon_e & \varepsilon_{max} & \theta_e & \Delta\theta & c_4(r_e) & {c_4^w}~~ & \delta\mu_1 & \delta\mu_2 \\
\noalign{\smallskip}
\hline\noalign{\smallskip}
{\rm a1}  &{\rm D33}  &{\rm E}    &00\ 56\ 31.5&-01\ 21\ 41&0.0418  &16.18&0.07&    & 0.41 &   4.03 &   0.58  &   20.83 & 0.20    & 10.00 &     4.46 & 1 &     0.05 &     0.18 &     32.6 &   66.8 &     0.23  \ \ &    0.04\ \ &     1.40 &     0.76\\
{\rm a2}  &{\rm D32}  &{\rm S0/E} &00\ 56\ 38.5&-01\ 21\ 07&        &16.76&0.03&    & 0.28 &   1.26 &   0.06  &   19.06 & 0.07    &  3.30 &     0.58 & 1 &     0.15 &     0.45 &     47.1 &    6.5 &        \ \ &   2.29\ \ &     1.48 &     0.41\\
{\rm a3}  &           &{\rm Sa/0} &00\ 57\ 02.9&-01\ 20\ 41&        &17.49&0.02&    & 0.18 &   2.60 &   0.06  &   21.33 & 0.03    &  1.90 &     0.20 & 1 &     0.48 &     0.54 &    123.7 &    8.9 &      2.08  \ \ &      -0.3  \ \ &     1.06 &     0.55\\
{\rm a4}  &           &{\rm Sa/0} &00\ 56\ 37.7&-01\ 20\ 41&        &18.27&0.03&    & 0.29 &   1.45 &   0.04  &   20.84 & 0.05    &  1.00 &     0.27 & 1 &     0.35 &     0.49 &     38.6 &    2.3 &    -0.19  \ \ &      2.08\ \ &     0.90 &     0.18\\
{\rm a5}  &           &{\rm dE}   &00\ 57\ 05.4&-01\ 19\ 20&        &18.68&0.08&    & 0.13 &   1.00 &   0.07  &   20.65 & 0.19    &  1.40 &     2.79 & 0 &     0.07 &     0.10 &     83.0 &   62.4 &         \ \ & -0.02\ \ &     0.93 &     0.18\\
\end{array}
      \]
   \end{table}
\end{landscape}

\begin{landscape}
\addtocounter{table}{-1}
   \begin{table}[p]
      \caption[]{{\bf b} -- Photometric and morphological parameters of galaxies in Abell~119(continue)}
         \label{GalPhot}
      \[
         \begin{array}{rcccccccccrrcrcccccrrrrcc}

\noalign{\bigskip}\hline\noalign{\smallskip}
{\rm Galaxy} & {\rm Name} & {\rm Type} & \alpha (2000) & \delta (2000) & z & {\rm m}_{_T} & \Delta{\rm m} & (B-r) & (V-r) & {\rm R}_{50} & \Delta{\rm R}_{50} & \50mu & \Delta{\rm \mu_{50}} & n & \Delta{n} & Q& \varepsilon_e & \varepsilon_{max} & \theta_e & \Delta\theta & c_4(r_e) & {c_4^w}~~ & \delta\mu_1 & \delta\mu_2 \\
\noalign{\smallskip}
\hline\noalign{\smallskip}
{\rm e2}  &{\rm D109} &{\rm S0}   &00\ 55\ 53.6&-00\ 54\ 03&0.0365  &16.13&0.04&1.22&      &   1.42 &   0.10  &   18.91 & 0.19    &  2.40 &     9.41 & 0 &     0.06 &     0.57 &     75.8 &   15.3 &    0.16  \ \ &     2.94 \ \ &     1.76 &     0.76\\
{\rm f1}  &           &{\rm S0}   &00\ 55\ 40.8&-00\ 59\ 47&        &17.36&0.01&    & 0.31 &   1.94 &   0.03  &   20.51 & 0.02    &  1.70 &     0.15 & 1 &     0.38 &     0.52 &    169.7 &    1.7 &         \ \ &   2.04 \ \ &     1.25 &     0.41\\
{\rm f2}  &{\rm D95}  &{\rm E}    &00\ 56\ 23.5&-00\ 59\ 12&0.0408  &15.30&0.03&    & 0.21 &   3.90 &   0.23  &   20.08 & 0.08    &  8.10 &     0.80 & 1 &     0.30 &     0.38 &     23.9 &    8.6 &   -0.30  \ \ &  -0.76 \ \ &     1.43 &     0.64\\
{\rm f3}  &           &{\rm S0}   &00\ 56\ 11.4&-00\ 58\ 13&        &17.52&0.03&    & 0.20 &   1.87 &   0.08  &   20.72 & 0.08    &  0.70 &     0.26 & 0 &     0.15 &     0.27 &     11.0 &    4.7 &    0.25  \ \ &   -0.32 \ \ &     1.14 &     0.86\\
{\rm f4}  &           &{\rm E}    &00\ 56\ 08.7&-00\ 57\ 54&0.0478  &16.89&0.09&    & 0.33 &   3.52 &   0.61  &   21.43 & 0.25    & 10.00 &     5.81 & 0 &     0.15 &     0.22 &     58.5 &    7.5 &  -0.09  \ \ &    -0.96 \ \ &     0.65 &     1.10\\
\end{array}
      \]
   \end{table}
\end{landscape}

\begin{landscape}
\addtocounter{table}{-1}
   \begin{table}[p]
      \caption[]{{\bf c} -- Photometric and morphological parameters of galaxies in Abell~1983}
         \label{GalPhot}
      \[
         \begin{array}{rcccccccccrrcrcccccrrrrcc}

\noalign{\bigskip}\hline\noalign{\smallskip}
{\rm Galaxy} & {\rm Name} & {\rm Type} & \alpha (2000) & \delta (2000) & z & {\rm m}_{_T} & \Delta{\rm m} & (B-r) & (V-r) & {\rm R}_{50} & \Delta{\rm R}_{50} & \50mu & \Delta{\rm \mu_{50}} & n & \Delta{n} & Q& \varepsilon_e & \varepsilon_{max} & \theta_e & \Delta\theta & c_4(r_e) & {c_4^w}~~ & \delta\mu_1 & \delta\mu_2 \\
\noalign{\smallskip}
\hline\noalign{\smallskip}
 {\rm a1} &{\rm D10}  &{\rm E}    &14\ 54\ 40.7& 16\ 13\ 18& 0.0434 &14.65&0.04&1.60&      &   3.79 &    0.30 &   19.58 &  0.125  &   4.60 &     1.58  & 1&     0.02 &     0.12 &   150.0 &  103.1 &   0.21   \ \ &   0.40  \ \ &     1.29 &     0.71\\
 {\rm b1} &           &{\rm Sa/0} &14\ 54\ 26.1& 16\ 20\ 13&        &16.86&0.02&1.19&      &   3.14 &    0.09 &   21.19 &  0.054  &   1.30 &     0.24  & 1&     0.34 &     0.37 &   144.2 &   15.8 &    2.37   \ \ &    1.15  \ \ &     1.11 &     0.46\\
 {\rm b2} &           &{\rm E}    &14\ 54\ 22.6& 16\ 20\ 48&        &16.56&0.02&1.35&      &   1.05 &    0.05 &   18.55 &  0.183  &   2.10 &     4.45  & 0&     0.14 &     0.14 &   136.8 &   74.0 &   0.33   \ \ &   0.33  \ \ &     1.86 &     1.13\\
 {\rm b3} &           &{\rm E}    &14\ 54\ 25.9& 16\ 21\ 07&        &16.11&0.03&1.34&      &   2.01 &    0.09 &   19.64 &  0.072  &   2.70 &     0.75  & 1&     0.05 &     0.13 &    50.4 &   26.4 &  -0.35   \ \ &   -0.42  \ \ &     1.34 &     0.76\\
 {\rm b4} &{\rm I4516}&{\rm E}    &14\ 54\ 23.4& 16\ 21\ 19& 0.0454 &12.46&0.05&1.36&      &  21.31 &    1.73 &   21.03 &  0.114  &   5.70 &     1.23  & 1&     0.26 &     0.33 &    26.5 &    5.8 &   0.32   \ \ &   0.23  \ \ &     1.29 &     0.66\\
\end{array}
      \]
   \end{table}
\end{landscape}

\begin{landscape}
\addtocounter{table}{-1}
   \begin{table}[p]
      \caption[]{{\bf d} -- Photometric and morphological parameters of galaxies in DC~2103}
         \label{GalPhot}
      \[
         \begin{array}{rcccccccccrrcrcccccrrrrcc}

\noalign{\bigskip}\hline\noalign{\smallskip}
{\rm Galaxy} & {\rm Name} & {\rm Type} & \alpha (2000) & \delta (2000) & z & {\rm m}_{_T} & \Delta{\rm m} & (B-r) & (V-r) & {\rm R}_{50} & \Delta{\rm R}_{50} & \50mu & \Delta{\rm \mu_{50}} & n & \Delta{n} & Q& \varepsilon_e & \varepsilon_{max} & \theta_e & \Delta\theta & c_4(r_e) & {c_4^w}~~ & \delta\mu_1 & \delta\mu_2 \\
\noalign{\smallskip}
\hline\noalign{\smallskip}
 {\rm a1} &           &{\rm Sa/0} &21\ 08\ 29.8&-39\ 55\ 47&        &18.51&0.06&    & -0.21&   1.64 &    0.14 &   21.89 &  0.11   &  10.00 &     1.31 & 0 &     0.32 &     0.36 &   117.7 &     6.2 &        \ \ &  0.57   \ \ &     0.99 &     0.64\\
 {\rm a2} &           &{\rm E}    &21\ 09\ 27.9&-39\ 55\ 22&        &17.23&0.04&    &  0.29&   3.91 &    0.24 &   22.25 &  0.09   &   3.70 &     0.75 & 1 &     0.10 &     0.23 &   161.2 &    16.5 &   0.77 \ \ &    0.92 \ \ &     1.37 &     0.52\\
 {\rm a3} &           &{\rm E}    &21\ 08\ 28.7&-39\ 55\ 08&        &18.36&0.33&    & -0.36&   4.42 &    0.67 &   23.39 &  0.26   &   9.80 &    20.00 & 0 &     0.23 &     0.28 &   106.8 &    38.3 &   0.41  \ \ &    0.57 \ \ &     1.30 &     1.09\\
 {\rm a4} &           &{\rm E}    &21\ 08\ 29.7&-39\ 54\ 59&        &16.02&0.02&    & -0.07&   3.17 &    0.09 &   20.75 &  0.04   &   2.40 &     0.33 & 1 &     0.16 &     0.17 &    33.4 &   125.1 &   0.36  \ \ &    0.15 \ \ &     1.19 &     0.65\\
 {\rm a5} &           &{\rm E}    &21\ 08\ 41.2&-39\ 53\ 09&        &17.92&0.01&    &  0.17&   8.32 &    0.16 &   24.31 &  0.04   &   9.90 &     1.64 & 0 &     0.36 &     0.41 &   145.6 &   160.8 &   0.05    \ \ &   0.02   \ \ &     1.17 &     0.71\\
\end{array}
      \]
   \end{table}
\end{landscape}

\begin{landscape}
\addtocounter{table}{-1}
   \begin{table}[p]
      \caption[]{{\bf e} -- Photometric and morphological parameters of galaxies in Abell~3125}
         \label{GalPhot}
      \[
         \begin{array}{rcccccccccrrcrcccccrrrrcc}

\noalign{\bigskip}\hline\noalign{\smallskip}
{\rm Galaxy} & {\rm Name} & {\rm Type} & \alpha (2000) & \delta (2000) & z & {\rm m}_{_T} & \Delta{\rm m} & (B-r) & (V-r) & {\rm R}_{50} & \Delta{\rm R}_{50} & \50mu & \Delta{\rm \mu_{50}} & n & \Delta{n} & Q& \varepsilon_e & \varepsilon_{max} & \theta_e & \Delta\theta & c_4(r_e) & {c_4^w}~~ & \delta\mu_1 & \delta\mu_2 \\
\noalign{\smallskip}
\hline\noalign{\smallskip}
 {\rm a1} &{\rm D9}   &{\rm E}    &03\ 26\ 53.8&-53\ 54\ 08&0.0615^*&15.58&0.02&    & 0.24 &   4.43 &    0.19 &   20.85 &  0.068  &   4.60 &     0.56 & 1 &     0.03 &     0.21 &  133.5 &    53.5 &   -0.28 \ \ &   -0.33 \ \ &     1.30 &     0.63\\
 {\rm a2} &           &{\rm E}    &03\ 26\ 19.8&-53\ 47\ 47&        &16.21&0.03&    & 0.32 &   8.29 &    0.46 &   22.79 &  0.072  &   6.60 &     0.71 & 1 &     0.27 &     0.34 &   28.8 &    16.5 &    0.62 \ \ &     1.15 \ \ &     1.22 &     0.73\\
 {\rm b1} &           &{\rm E}    &03\ 29\ 51.8&-53\ 34\ 17&        &17.17&0.03&    & 0.20 &   2.95 &    0.14 &   21.77 &  0.070  &   2.70 &     0.47 & 1 &     0.08 &     0.19 &    0.4 &    41.8 &    0.58 \ \ & -0.01 \ \ &     1.07 &     0.54\\
 {\rm b2} &           &{\rm E}    &03\ 29\ 51.8&-53\ 34\ 07&        &17.63&0.01&    & 0.22 &   1.31 &    0.03 &   20.21 &  0.030  &   7.30 &     0.77 & 0 &     0.19 &     0.27 &  142.6 &    11.7 &        \ \ &    0.55 \ \ &     1.51 &     0.84\\
 {\rm b3} &           &{\rm S0}   &03\ 28\ 54.7&-53\ 29\ 15&        &18.06&0.03&    & 0.15 &   1.49 &    0.08 &   20.99 &  0.097  &   1.10 &     0.42 & 0 &     0.15 &     0.25 &   51.3 &     9.0 &        \ \ &    -0.30 \ \ &     1.63 &     0.57\\
\end{array}
      \]
   \end{table}
\end{landscape}

\begin{landscape}
\addtocounter{table}{-1}
   \begin{table}[p]
      \caption[]{{\bf e} -- Photometric and morphological parameters of galaxies in Abell~3125(continue)}
         \label{GalPhot}
      \[
         \begin{array}{rcccccccccrrcrcccccrrrrcc}

\noalign{\bigskip}\hline\noalign{\smallskip}
{\rm Galaxy} & {\rm Name} & {\rm Type} & \alpha (2000) & \delta (2000) & z & {\rm m}_{_T} & \Delta{\rm m} & (B-r) & (V-r) & {\rm R}_{50} & \Delta{\rm R}_{50} & \50mu & \Delta{\rm \mu_{50}} & n & \Delta{n} & Q& \varepsilon_e & \varepsilon_{max} & \theta_e & \Delta\theta & c_4(r_e) & {c_4^w}~~ & \delta\mu_1 & \delta\mu_2 \\
\noalign{\smallskip}
\hline\noalign{\smallskip}
 {\rm f1} &           &{\rm Sa/0} &03\ 24\ 10.0&-53\ 11\ 28&        &18.09&0.02&    & 0.16 &   1.87 &    0.05 &   21.48 &  0.043  &   1.00 &     0.18 & 0 &     0.17 &     0.33 &  133.4 &     4.3 &         \ \ &    2.50  \ \ &     0.77 &     0.72\\
 {\rm f2} &           &{\rm S0/a} &03\ 24\ 01.7&-53\ 08\ 18&        &17.92&0.02&    & 0.24 &   1.57 &    0.03 &   20.98 &  0.030  &   1.10 &     0.15 & 1 &     0.33 &     0.50 &   99.3 &     3.2 &         \ \ &    1.78  \ \ &     0.97 &     0.68\\
 {\rm f3} &{\rm D97}  &{\rm E/S0} &03\ 24\ 01.7&-53\ 08\ 18&        &16.33&0.02&    & 0.13 &   2.48 &    0.06 &   20.35 &  0.042  &   1.20 &     0.20 & 1 &     0.23 &     0.32 &  116.7 &     3.8 &   0.54   \ \ &   -0.32  \ \ &     1.08 &     0.73\\
 {\rm f4} &           &{\rm S0}   &03\ 24\ 33.7&-53\ 07\ 42&        &17.16&0.04&    & 0.24 &   2.29 &    0.16 &   21.00 &  0.113  &   1.00 &     0.41 & 0 &     0.06 &     0.11 &  137.5 &    44.4 &   -1.47   \ \ &   0.44  \ \ &     1.29 &     0.90\\
 {\rm f5} &           &{\rm E}    &03\ 24\ 11.6&-53\ 06\ 56&        &16.91&0.04&    & 0.18 &   1.71 &    0.10 &   20.13 &  0.100  &   2.00 &     0.76 & 1 &     0.22 &     0.27 &  146.0 &    33.3 &         \ \ & -0.04  \ \ &     1.63 &     0.49\\
\end{array}
      \]
   \end{table}
\end{landscape}

\begin{landscape}
\addtocounter{table}{-1}
   \begin{table}[p]
      \caption[]{{\bf f} -- Photometric and morphological parameters of galaxies in Abell~1069}
         \label{GalPhot}
      \[
         \begin{array}{rcccccccccrrcrcccccrrrrcc}

\noalign{\bigskip}\hline\noalign{\smallskip}
{\rm Galaxy} & {\rm Name} & {\rm Type} & \alpha (2000) & \delta (2000) & z & {\rm m}_{_T} & \Delta{\rm m} & (B-r) & (V-r) & {\rm R}_{50} & \Delta{\rm R}_{50} & \50mu & \Delta{\rm \mu_{50}} & n & \Delta{n} & Q& \varepsilon_e & \varepsilon_{max} & \theta_e & \Delta\theta & c_4(r_e) & {c_4^w}~~ & \delta\mu_1 & \delta\mu_2 \\
\noalign{\smallskip}
\hline\noalign{\smallskip}
 {\rm a1} &           &{\rm Sa/0} &10\ 37\ 23.1&-09\ 16\ 10&        &19.43&0.05&1.38&      &   2.13 &    0.15 &   23.04 &  0.18   &  1.00 &     0.94  & 0&     0.44 &     0.47 &   53.3 &     6.4 &   2.47  \ \ &     1.37 \ \ &     1.39 &     0.28\\
 {\rm a2} &           &{\rm E}    &10\ 37\ 53.7&-09\ 09\ 05&        &17.29&0.04&1.31&      &   1.59 &    0.09 &   20.31 &  0.11   &  1.50 &     0.74  & 1&     0.17 &     0.25 &   20.0 &    14.8 &  0.87  \ \ &  -0.02 \ \ &     1.53 &     0.70\\
 {\rm a3} &{\rm D3}   &{\rm S0}   &10\ 37\ 31.7&-09\ 08\ 56&0.0636^*&16.38&0.12&1.42&      &   1.84 &    0.42 &   19.75 &  0.33   &  2.00 &     8.87  & 0&     0.26 &     0.59 &   39.8 &     1.6 &   1.83  \ \ &     4.26 \ \ &     1.64 &     0.63\\
 {\rm a4} &           &{\rm Sa/0} &10\ 37\ 48.7&-09\ 07\ 53&        &19.49&0.04&1.29&      &   1.83 &    0.11 &   22.81 &  0.12   &  1.50 &     0.82  & 1&     0.29 &     0.35 &   93.5 &    10.6 &   2.32  \ \ &      1.81 \ \ &     1.48 &     0.39\\
 {\rm a5} &           &{\rm S0/a} &10\ 37\ 22.7&-09\ 03\ 23&        &18.70&0.03&1.36&      &   1.28 &    0.03 &   21.32 &  0.05   &  0.90 &     0.23  & 1&     0.39 &     0.43 &   71.7 &     3.1 &       \ \ &    0.97 \ \ &     0.92 &     0.20\\
\end{array}
      \]
   \end{table}
\end{landscape}
\begin{landscape}
\addtocounter{table}{-1}
   \begin{table}[p]
      \caption[]{{\bf f} -- Photometric and morphological parameters of galaxies in Abell~1069(continue)}
         \label{GalPhot}
      \[
         \begin{array}{rcccccccccrrcrcccccrrrrcc}

\noalign{\bigskip}\hline\noalign{\smallskip}
{\rm Galaxy} & {\rm Name} & {\rm Type} & \alpha (2000) & \delta (2000) & z & {\rm m}_{_T} & \Delta{\rm m} & (B-r) & (V-r) & {\rm R}_{50} & \Delta{\rm R}_{50} & \50mu & \Delta{\rm \mu_{50}} & n & \Delta{n} & Q& \varepsilon_e & \varepsilon_{max} & \theta_e & \Delta\theta & c_4(r_e) & {c_4^w}~~ & \delta\mu_1 & \delta\mu_2 \\
\noalign{\smallskip}
\hline\noalign{\smallskip}
{\rm f12} &           &{\rm S0}   &10\ 38\ 12.7&-08\ 06\ 00&        &18.97&0.05&    &  0.26&   1.08 &    0.09 &   21.17 &  0.16   &  1.40 &     1.03  & 1&     0.28 &     0.36 &  172.3 &     2.6 &          \ \ &   1.31  \ \ &     1.60 &     0.63\\
 {\rm g1} &           &{\rm E}    &10\ 40\ 59.6&-08\ 03\ 15&        &17.53&0.08&    &  0.26&   1.02 &    0.14 &   19.65 &  0.25   &  2.40 &     2.90  & 0&     0.16 &     0.23 &  105.7 &    35.8 &          \ \ &  0.67  \ \ &     1.71 &     0.60\\
 {\rm g2} &           &{\rm E/S0} &10\ 40\ 29.2&-08\ 02\ 11&        &18.54&0.05&    &  0.35&   1.49 &    0.10 &   21.46 &  0.11   &  1.90 &     0.73  & 1&     0.24 &     0.31 &  149.8 &     7.4 &     0.54  \ \ & -0.12  \ \ &     1.41 &     0.69\\
 {\rm g3} &           &{\rm E/S0} &10\ 40\ 23.5&-08\ 01\ 50&        &18.15&0.06&    &  0.32&   1.30 &    0.15 &   20.81 &  0.21   &  1.50 &     1.35  & 0&     0.12 &     0.19 &   52.4 &     9.7 &          \ \ & -0.46  \ \ &     1.75 &     0.71\\
 {\rm g4} &           &{\rm E/S0} &10\ 40\ 34.5&-08\ 00\ 52&        &16.19&0.06&    &  0.30&   2.28 &    0.27 &   19.97 &  0.19   &  3.10 &     1.89  & 1&     0.26 &     0.39 &   60.1 &     8.5 &      1.43  \ \ &  0.27  \ \ &     1.32 &     1.14\\
\end{array}
      \]
   \end{table}
\end{landscape}

\begin{landscape}
\addtocounter{table}{-1}
   \begin{table}[p]
      \caption[]{{\bf g} -- Photometric and morphological parameters of galaxies in Abell~2670}
         \label{GalPhot}
      \[
         \begin{array}{rcccccccccrrcrcccccrrrrcc}

\noalign{\bigskip}\hline\noalign{\smallskip}
{\rm Galaxy} & {\rm Name} & {\rm Type} & \alpha (2000) & \delta (2000) & z & {\rm m}_{_T} & \Delta{\rm m} & (B-r) & (V-r) & {\rm R}_{50} & \Delta{\rm R}_{50} & \50mu & \Delta{\rm \mu_{50}} & n & \Delta{n} & Q& \varepsilon_e & \varepsilon_{max} & \theta_e & \Delta\theta & c_4(r_e) & {c_4^w}~~ & \delta\mu_1 & \delta\mu_2 \\
\noalign{\smallskip}
\hline\noalign{\smallskip}
 {\rm a1} &           &    {\rm E}&23\ 54\ 10.1&-10\ 27\ 50&        &17.44&0.06&1.48&      &  1.12  &    0.11 &   19.81 &  0.18   &   1.60 &     1.19 & 0 &     0.08 &     0.18 &   131.3 &    39.4 &   0.21  \ \ &   -0.64 \ \ &     1.52 &     0.85\\
 {\rm a2} &           & {\rm E/S0}&23\ 54\ 16.5&-10\ 27\ 43&        &18.10&0.01&1.40&      &  1.07  &    0.02 &   20.28 &  0.02   &   1.30 &     0.08 & 0 &     0.06 &     0.15 &   160.2 &    71.6 &  -0.13  \ \ &  -0.05 \ \ &     0.83 &     1.16\\
 {\rm a3} &           &    {\rm E}&23\ 54\ 13.5&-10\ 27\ 18&        &18.06&0.02&1.22&      &  0.70  &    0.03 &   19.34 &  0.06   &   2.50 &     0.23 & 0 &     0.21 &     0.27 &   151.1 &    12.5 &    1.04  \ \ &    0.30 \ \ &     1.31 &     1.04\\
 {\rm a4} &           &    {\rm E}&23\ 54\ 10.7&-10\ 27\ 06& 0.0761 &18.33&0.02&1.39&      &  1.79  &    0.04 &   20.83 &  0.03   &   2.30 &     0.22 & 1 &     0.09 &     0.14 &   133.3 &    24.9 &    1.31  \ \ &   -0.18 \ \ &     0.91 &     0.53\\
 {\rm a5} &           &    {\rm E}&23\ 54\ 09.0&-10\ 26\ 36&        &18.28&0.11&1.38&      &  0.45  &    0.10 &   18.61 &  0.30   &   3.20 &    11.28 & 0 &     0.14 &     0.17 &   178.8 &    72.4 &        \ \ &    0.35 \ \ &     1.72 &     0.40\\
\end{array}
      \]
   \end{table}
\end{landscape}

\end{changemargin}


\newpage
   \begin{figure*}[p]
      \vspace{2.5cm}
      \vspace{1.5cm}
      \caption{
(a) -- Overlap between our CCD fields and DSS imaging for Abell~2151.
In Table~\ref{Fields}
the fields are listed in ascending order of declination. The galaxies
belonging to our sample in each field are marked with small
circles. The absolute positions are set by the coordinates (Right Ascension
[2000] and Declination [2000])
corresponding to the cross in the plots, while the grid sizes in
$\alpha$ and $\delta$ are of $30^s$ and $5^\prime$, respectively. In
this figure and in the following ones, relative to the other clusters,
the colors of the fields refer to a given telescope+camera
configuration: (i) blue for run~\#1; (ii) green for runs~\#2 and \#3;
(iii) orange for runs~\#4 and \#5; (iv) red for runs~\#6 and \#7.
      }
      \label{Clu}
    \end{figure*}

\addtocounter{figure}{-1}

   \begin{figure*}[p]
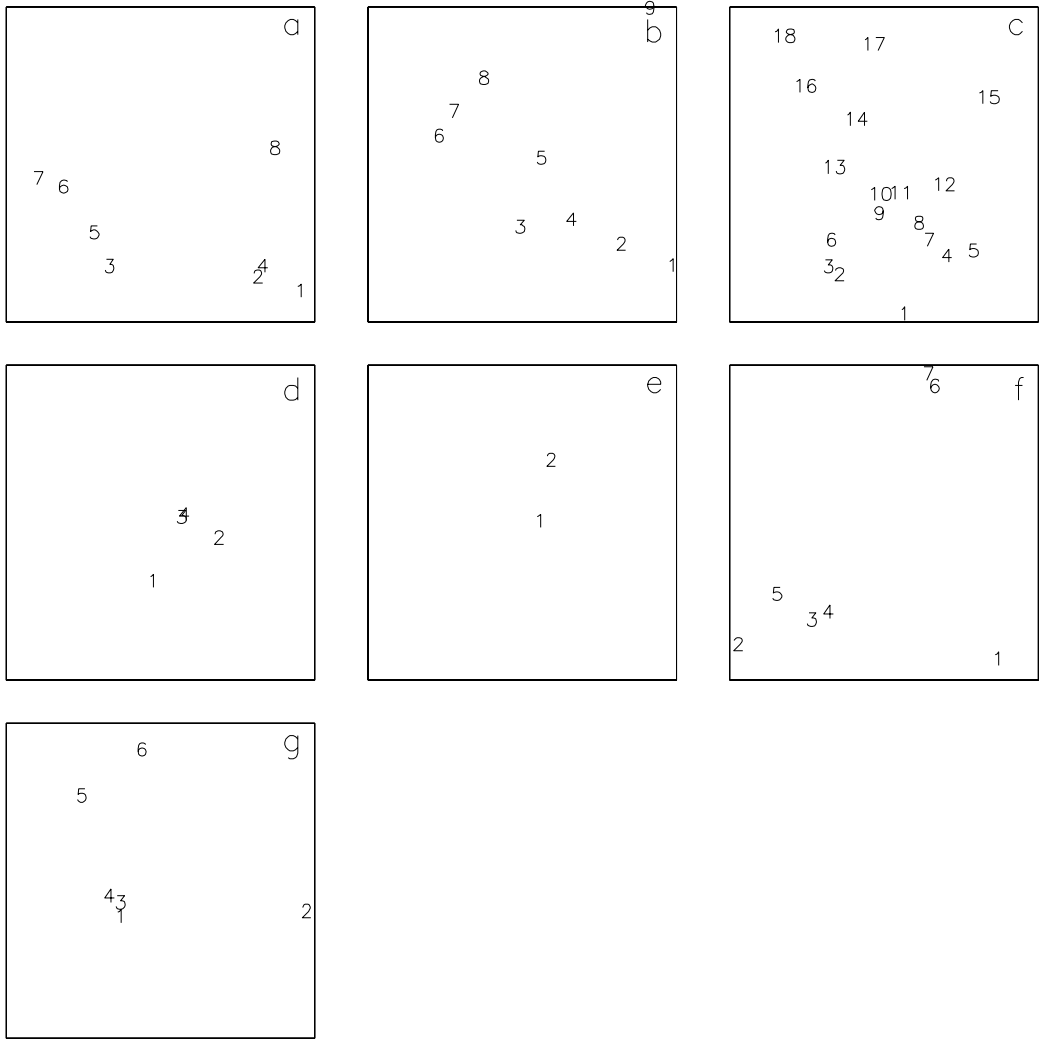

      \caption{
(b) -- Same as Figure~\ref{Clu}a, but for Abell~119.
      }
      \label{Clu}
    \end{figure*}

\addtocounter{figure}{-1}

   \begin{figure*}[p]
      \caption{
(c) -- Same as Figure~\ref{Clu}a, but for Abell~1983.
      }
      \label{Clu}
    \end{figure*}

\newpage
\addtocounter{figure}{-1}

   \begin{figure*}[p]
      \caption{
(d) -- Same as Figure~\ref{Clu}a, but for DC~2103.
      }
      \label{Clu}
    \end{figure*}

\newpage
\addtocounter{figure}{-1}

   \begin{figure*}[p]
      \caption{
(e) -- Same as Figure~\ref{Clu}a, but for Abell~3125.
      }
      \label{Clu}
    \end{figure*}

\newpage
\addtocounter{figure}{-1}

   \begin{figure*}[p]
      \caption{
(f) -- Same as Figure~\ref{Clu}a, but for Abell~1069.
      }
      \label{Clu}
    \end{figure*}

\newpage
\addtocounter{figure}{-1}

   \begin{figure*}[p]
      \caption{
(g) -- Same as Figure~\ref{Clu}a, but for Abell~2670.
      }
      \label{Clu}
    \end{figure*}

\clearpage
\newpage
   \begin{figure*}[p]
      \resizebox{\hsize}{!}{\includegraphics{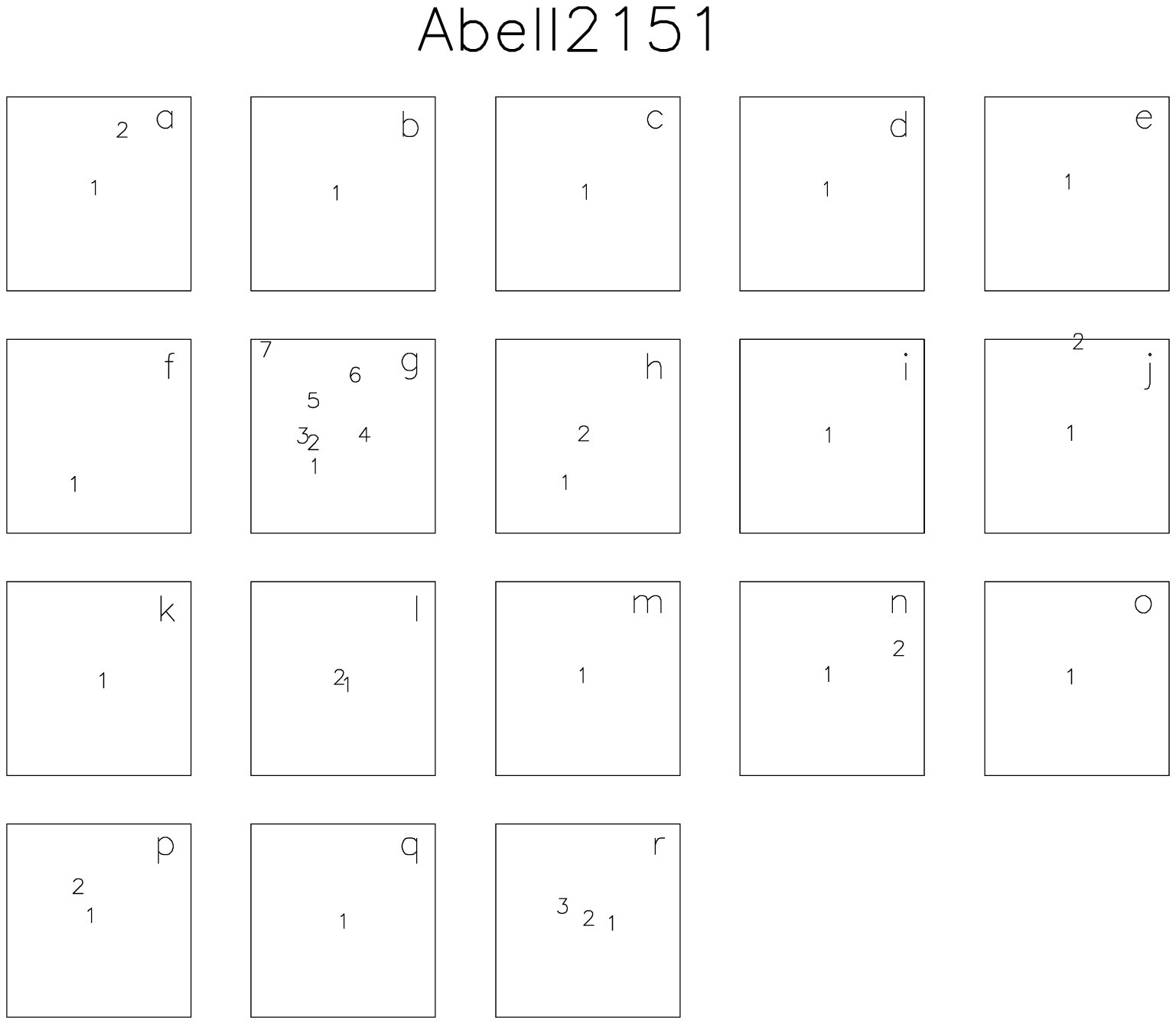}}
      \caption{
(a) -- Identification of the selected galaxies in our CCD fields of
Abell~2151. The galaxies are numbered in ascending order of
declination. The alphabetic order of the letters identifying the
fields corresponds to the ascending order of declination (see Table~\ref{Fields}
and caption of Figure~\ref{Clu}).
      }
      \label{Cluf}
    \end{figure*}

\newpage
\addtocounter{figure}{-1}

   \begin{figure*}[p]
      \resizebox{\hsize}{!}{\includegraphics{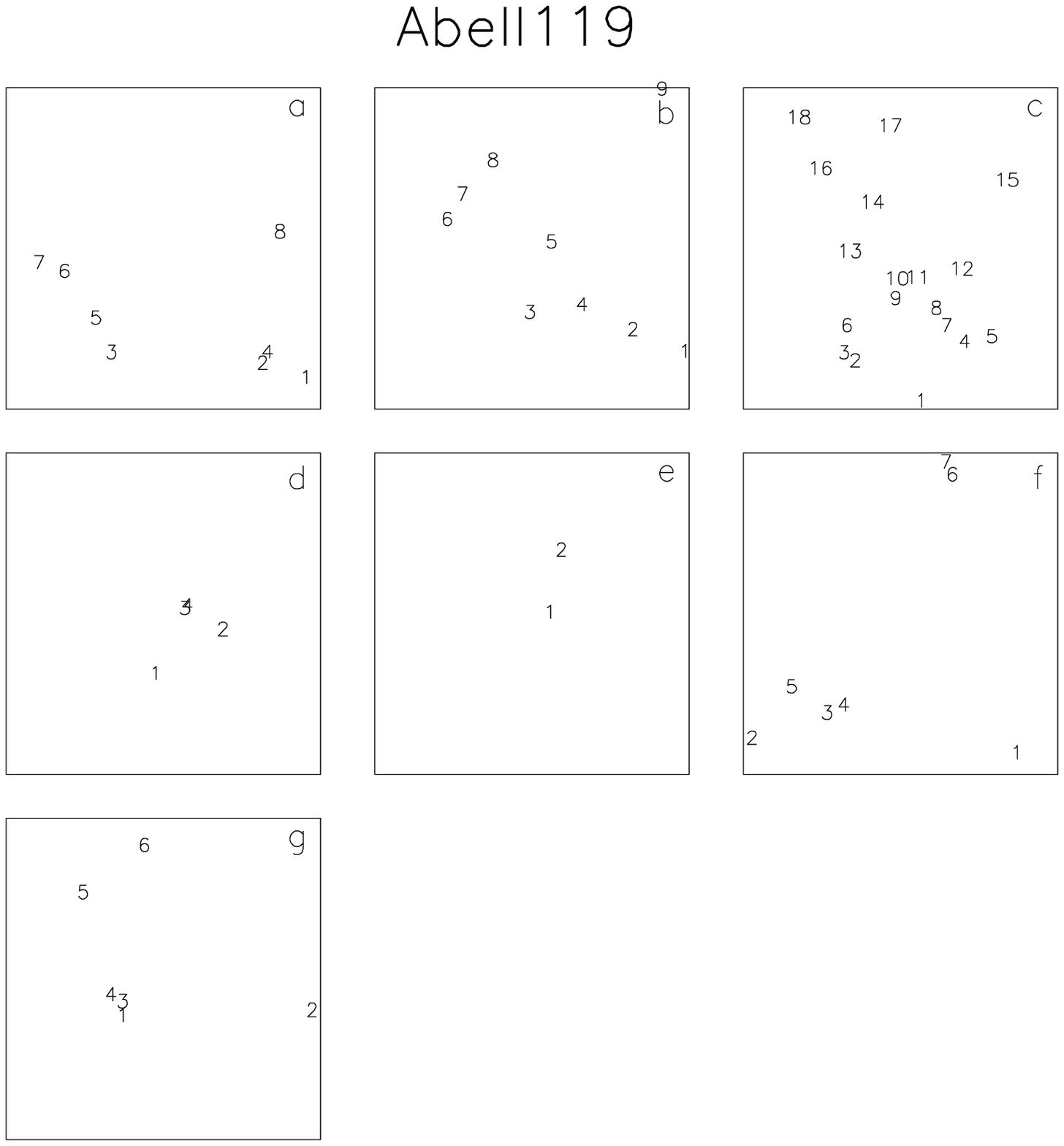}}
      \caption{
(b) -- Same as Figure~\ref{Cluf}a, but for Abell~119.
      }
      \label{Cluf}
    \end{figure*}

\newpage
\addtocounter{figure}{-1}

\clearpage
   \begin{figure*}[p]
      \resizebox{\hsize}{!}{\includegraphics{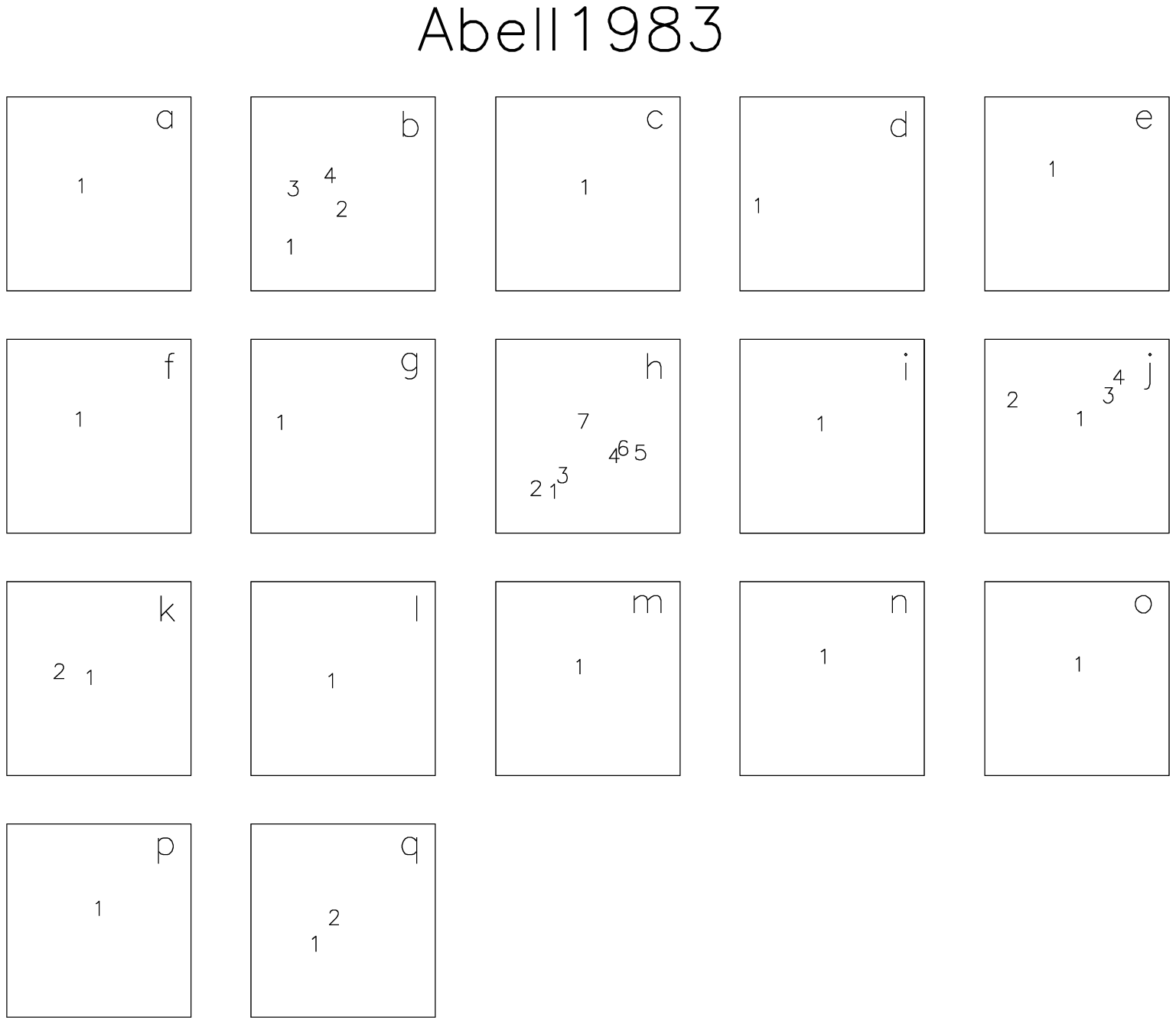}}
      \caption{
(c) -- Same as Figure~\ref{Cluf}a, but for Abell~1983.
      }
      \label{Cluf}
    \end{figure*}

\newpage
\addtocounter{figure}{-1}

   \begin{figure*}[p]
      \resizebox{\hsize}{!}{\includegraphics{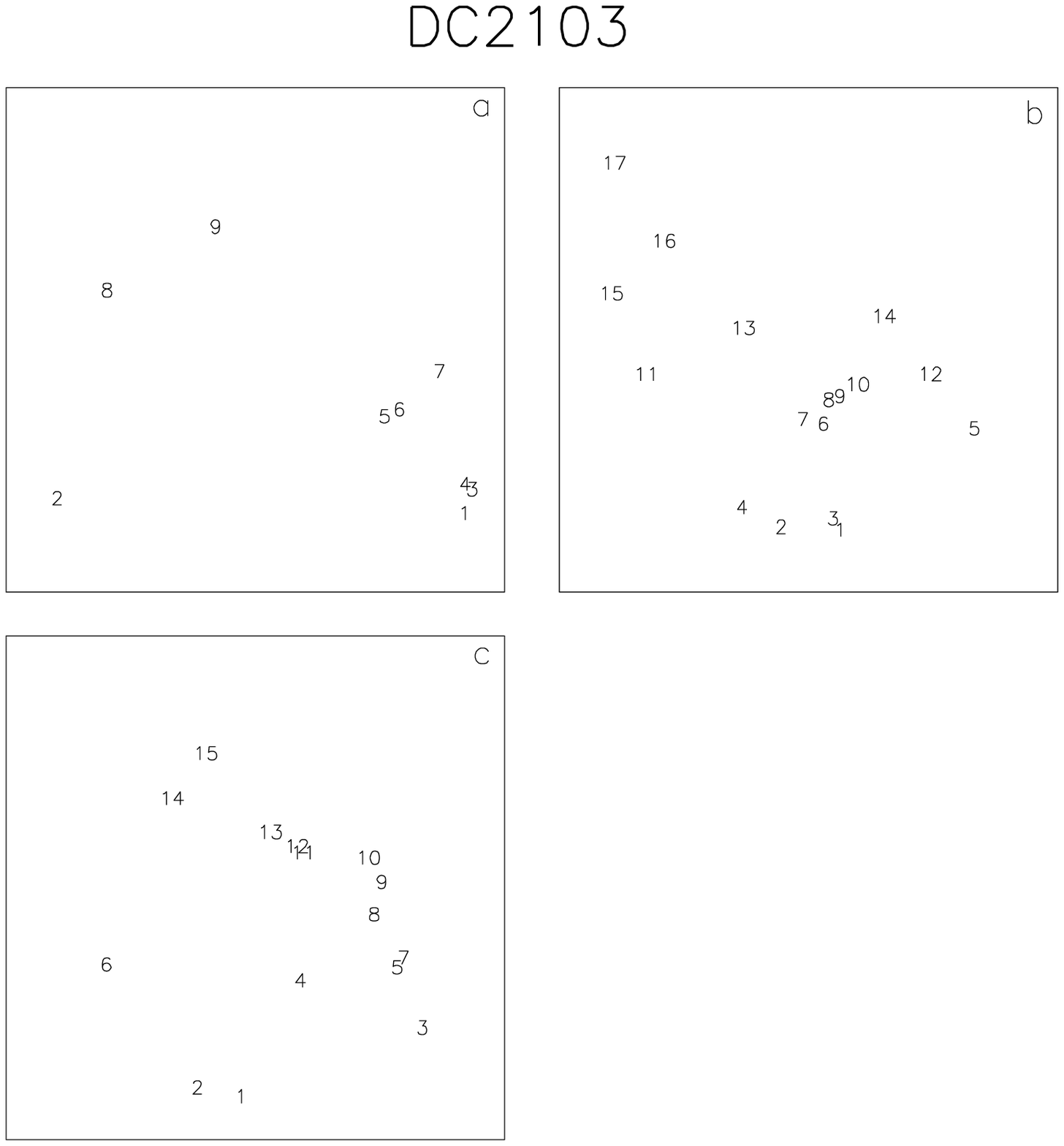}}
      \caption{
(d) -- Same as Figure~\ref{Cluf}a, but for DC~2103.
      }
      \label{Cluf}
    \end{figure*}

\newpage
\addtocounter{figure}{-1}

   \begin{figure*}[p]
      \resizebox{\hsize}{!}{\includegraphics{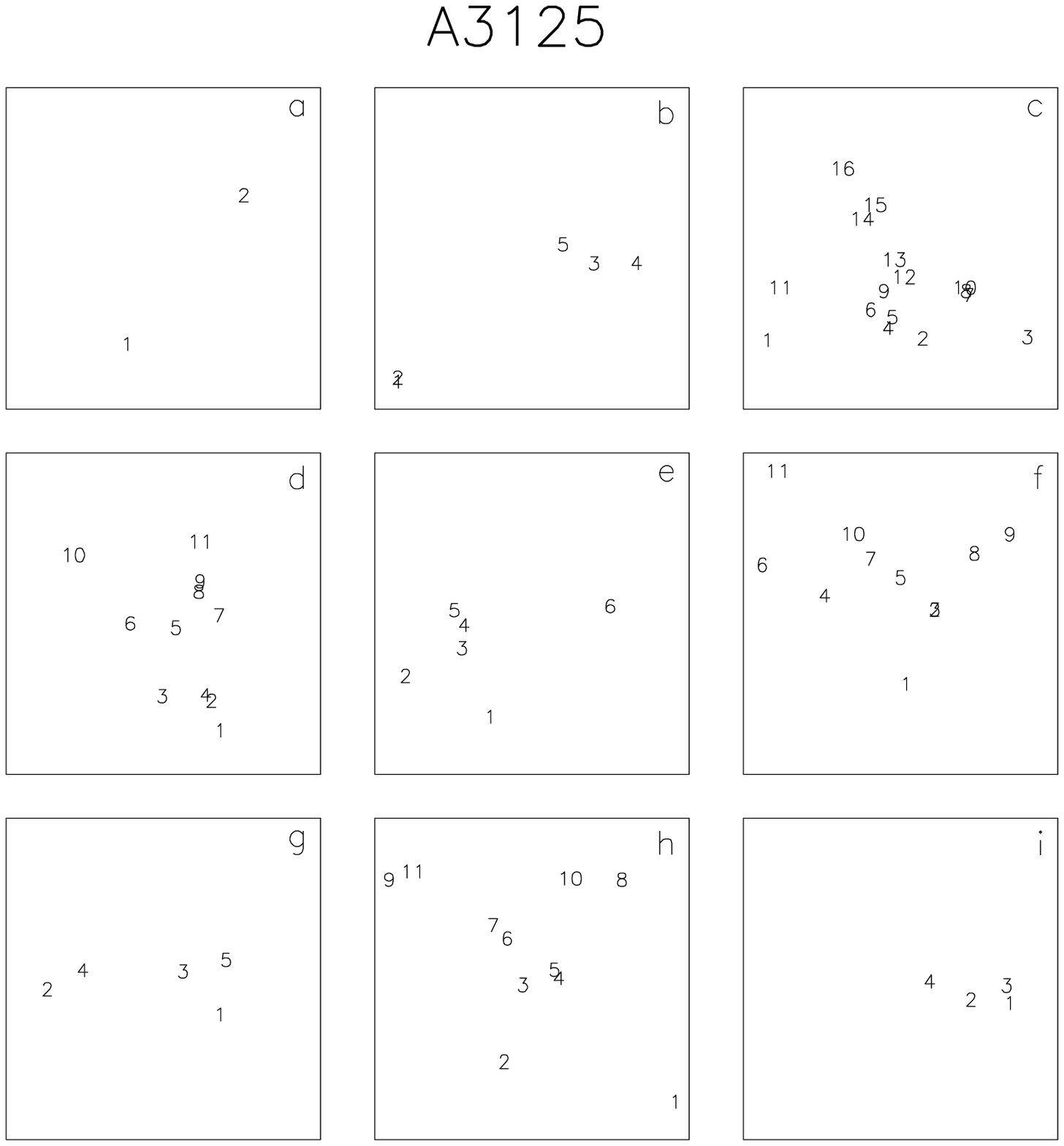}}
      \caption{
(e) -- Same as Figure~\ref{Cluf}a, but for Abell~3125.
      }
      \label{Cluf}
    \end{figure*}

\newpage
\addtocounter{figure}{-1}

   \begin{figure*}[p]
      \resizebox{\hsize}{!}{\includegraphics{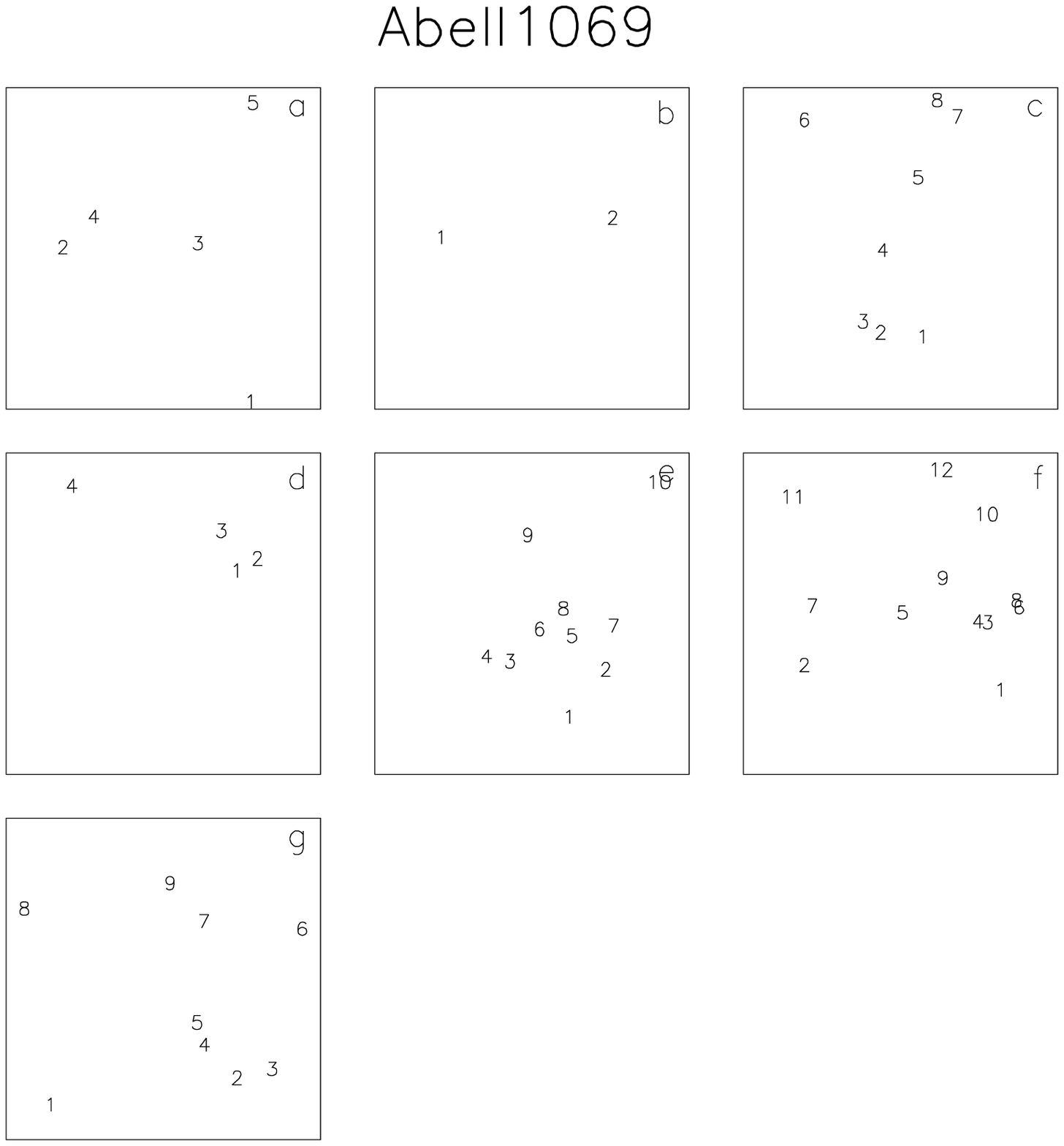}}
      \caption{
(f) -- Same as Figure~\ref{Cluf}a, but for Abell~1069.
      }
      \label{Cluf}
    \end{figure*}

\newpage
\addtocounter{figure}{-1}

   \begin{figure*}[p]
      \resizebox{\hsize}{!}{\includegraphics{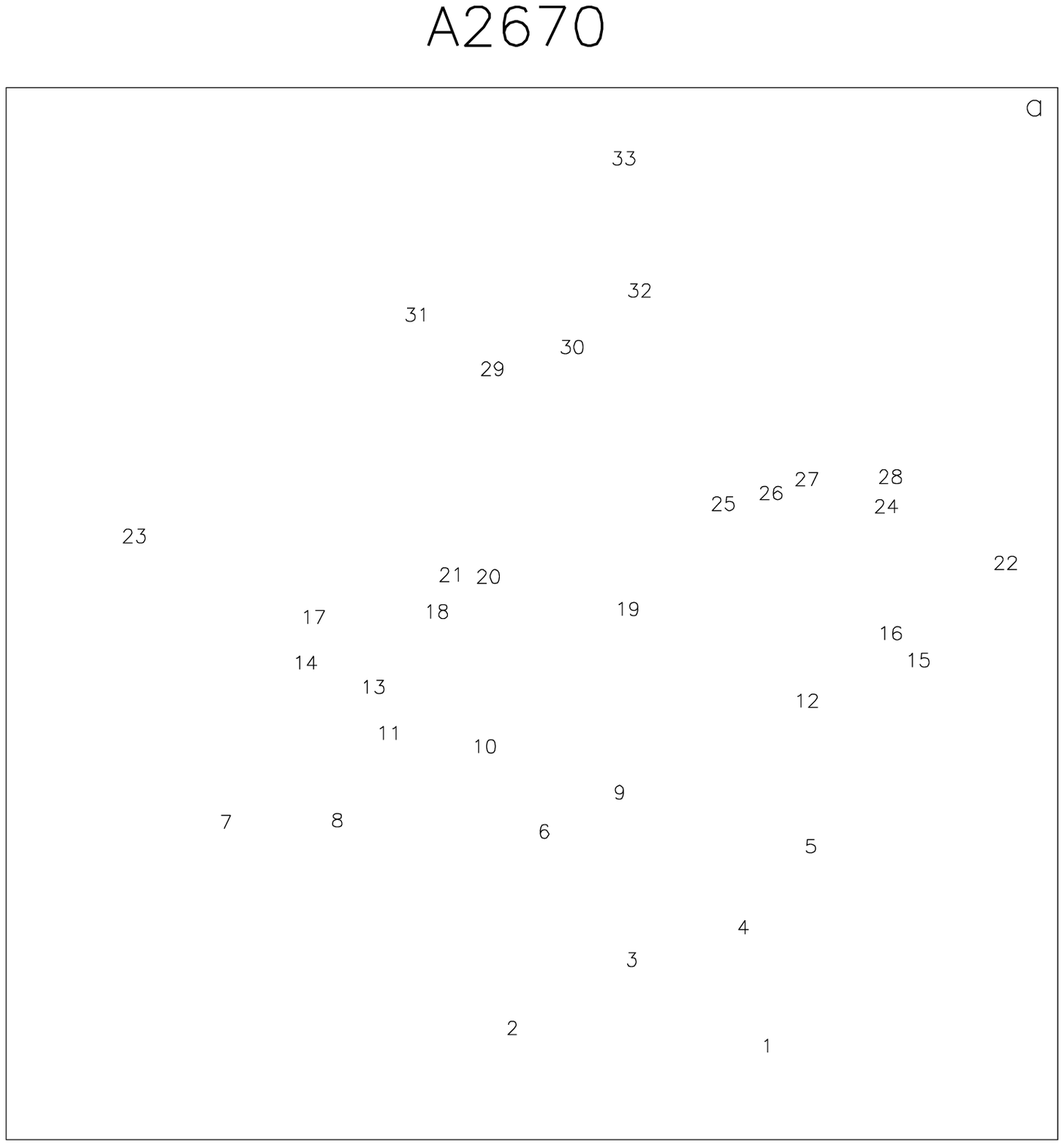}}
      \caption{
(g) -- Same as Figure~\ref{Cluf}a, but for Abell~2670.
      }
      \label{Cluf7}
    \end{figure*}

\end{document}